\documentclass[12pt,onecolumn,draftclsnofoot,]{IEEEtran}
\IEEEoverridecommandlockouts
\usepackage{graphicx} 
\usepackage{caption}
\usepackage{amsmath,amssymb,amsfonts}
\usepackage{algorithmic}
\usepackage{textcomp}
\usepackage{subcaption}
\usepackage{xcolor}
\usepackage[textsize=tiny]{todonotes}

\usepackage{tikz}
\usetikzlibrary{arrows,automata, positioning}
\usetikzlibrary{calc, chains,
                fit,
                positioning,
                shapes}
\usetikzlibrary{patterns}       

\usepackage{pgfplots}
\usepackage[english]{babel}

\usepackage{dashrule}

\usepackage{float}
\usepackage{comment}
\usepackage[acronym,shortcuts]{glossaries}

\usepackage{theorem}
{\theorembodyfont{\rmfamily}
   }

\newacronym{AWGN}{AWGN}{Additive White Gaussian Noise}
\newacronym{BCH}{BCH}{Bose–Chaudhuri–Hocquenghem}
\newacronym{CCSDS}{CCSDS}{Consultative Committee for Space Data Systems}
\newacronym{CER}{CER}{Codeword Error Rate}
\newacronym{BPSK}{BPSK}{Binary Phase-Shift Keying}
\newacronym{CLTU}{CLTU}{Communications Link Transmission Unit}
\newacronym{LDPC}{LDPC}{Low-Density Parity-Check}
\newacronym{LFSR}{LFSR}{Linear-Feedback Shift-Register}
\newacronym{LLR-SPA}{LLR-SPA}{Log-Likelihood Ratio Sum-Product Algorithm}
\newacronym{ML}{ML}{Maximum Likelihood}
\newacronym{MSA}{MSA}{Min-Sum Algorithm}
\newacronym{NMSA}{NMSA}{Normalized Min-Sum Algorithm}
\newacronym{S-LRT}{S-LRT}{Simplified Likelihood Ratio Test}
\newacronym{SNR}{SNR}{Signal-to-Noise Ratio}
\newacronym{SPA}{SPA}{Sum-Product Algorithm}
\newacronym{TC}{TC}{TeleCommand}
\newacronym{TS}{TS}{Tail Sequence}

\title{
Telecommand Rejection Probability for CCSDS-compliant LDPC-Coded Transmissions with Tail Sequence}

\author{
Rebecca Giuliani\textsuperscript{1}, Massimo Battaglioni\textsuperscript{1,2}, Marco Baldi\textsuperscript{1,2}, Franco Chiaraluce\textsuperscript{1,2}, and Nicola Maturo\textsuperscript{3}\\
\IEEEauthorblockA{\textsuperscript{1}\textit{Dipartimento di Ingegneria dell'Informazione, Università Politecnica delle Marche, Ancona (60131), Italy\\\textsuperscript{2}\textit{Consorzio Nazionale Interuniversitario per le Telecomunicazioni, 43124 Parma,
Italy }}\\
email: r.giuliani@pm.univpm.it, \{m.battaglioni, m.baldi, f.chiaraluce\}@univpm.it}
\IEEEauthorblockA{\textsuperscript{3}\textit{European Space Agency, ESTEC}, Keplerlaan 1, PO Box 299, NL-2200 AG Noordwijk, The Netherlands\\
email: nicola.maturo@esa.int}
}

\begin{document}

\maketitle

\begin{abstract}
According to the Consultative Committee for Space Data Systems (CCSDS) recommendation for TeleCommand (TC) synchronization and coding, the Communications Link Transmission Unit (CLTU) consists of a start sequence, followed by coded data, and a tail sequence, which might be optional depending on the employed coding scheme. 
With regard to the latter, these transmissions traditionally use a modified Bose–Chaudhuri–Hocquenghem (BCH) code, to which two state-of-the-art Low-Density Parity-Check (LDPC) codes were later added.
As a lightweight technique to detect the presence of the tail sequence, an approach based on decoding failure has traditionally been used, choosing a non-correctable string as the tail sequence.
This works very well with the BCH code, for which bounded-distance decoders are employed. When the same approach is employed with LDPC codes, it is necessary to design the tail sequence as a non-correctable string for the case of iterative decoders based on belief propagation.
Moreover, the tail sequence might be corrupted by noise, potentially converting it into a correctable pattern. 
It is therefore important that the tail sequence is chosen to be as much distant as possible, according to some metric, from any legitimate codeword.
In this paper we study such problem, and analyze the TC rejection probability both theoretically and through simulations. Such a performance figure, being the rate at which the CLTU is discarded, should clearly be minimized. Our analysis is performed considering many different choices of the system parameters (e.g., length of the CLTU, decoding algorithm, maximum number of decoding iterations).
\end{abstract}


\section{Introduction}

\begin{figure}[b]
    \centering
    \resizebox{0.8\textwidth}{!}{
\begin{tikzpicture}

\tikzstyle{box} = [rectangle, minimum height=1cm, minimum width=2cm, text centered, draw=black]
\tikzstyle{arrow} = [thick,->,>=stealth]

\node[box, fill=cyan!20] (start) {Start Sequence};
\node[box, right=0.1cm of start, fill=green!20] (cw1) {Codeword $1$};
\node[box, right=0.1cm of cw1, fill=green!20] (cw2) {$\cdots$};
\node[box, right=0.1cm of cw2, fill=green!20] (cw3) {Codeword $N$};
\node[box, right=0.1cm of cw3, fill=red!20] (ts) {Tail Sequence};

\node[below=0.5cm of start] {Synchronizes beginning of \ac{CLTU}};
\node[below=0.5cm of cw2] {Encoded data};
\node[below=0.5cm of ts] {Delimits end of \ac{CLTU}};

\end{tikzpicture}}
    \caption{CLTU structure \cite{bluebook}}
    \label{fig:CLTU}
\end{figure}
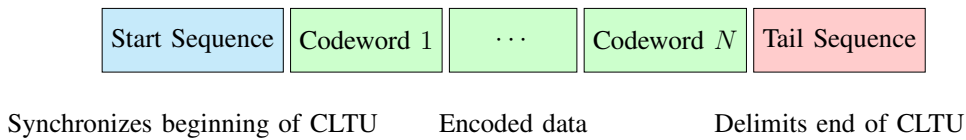

In space missions, the \ac{TC} function plays a crucial role, in that it is responsible for the transmission of commands to the spacecraft. 
The \ac{CCSDS} suggests that, in order to be reliably received and correctly processed by the space element, raw data need to be encoded and encapsulated into a \ac{CLTU}. According to \cite{bluebook}, and as described in \cite{tutorial}, the \ac{CLTU} should be formed, sequentially, by:
\begin{itemize}
    \item a start sequence, aimed at synchronizing the beginning of a \ac{CLTU} and at delimiting the beginning of the first codeword;
    \item a certain number of codewords, say $N$, representing the encoded data;
    \item a \ac{TS} delimiting the end of the \ac{CLTU} that is optional, depending on the employed error correcting coding scheme.
\end{itemize}
The structure of the \ac{CLTU} is shown in Fig. \ref{fig:CLTU}. Due to its role, the start sequence should be designed as a pattern with good autocorrelation properties, such that the use of a classical correlation-based detector yields negligible probability of confusing it with another pattern. When an \ac{AWGN} channel is considered, the optimal strategy to detect a periodically inserted noisy sync sequence is that of adopting the algorithm proposed in \cite{massey1972optimum}. If there is a single sync sequence, it is possible to use the \ac{S-LRT} \cite{Chiani2006, Pfletschinger2015}. 

As for data reliability, many error correcting coding schemes might be adopted to mitigate the effect of the noise introduced by the channel. Notable examples are \ac{BCH} codes and \ac{LDPC} codes, first introduced in the seminal works \cite{hocquenghem1959codes,bose1960class} and \cite{gallager1960high}, respectively. These families of codes are those recommended by the \ac{CCSDS} in \cite{bluebook}.

Differently from the start sequence, the design of the \ac{TS} is quite challenging, because it must take into account the technique used for its detection. Possible methods include using a hard/soft correlator, or the \ac{S-LRT}, similar to the approach for the start sequence. An engaging alternative is that of designing the \ac{TS} in such a way that it triggers an error in the decoding process. This way, the receiver does not need to switch between devices (decoder and correlator), and just continues decoding until it fails. It is important to note that each method imposes distinct requirements. For example, the former approaches generally require low side lobes in the auto-correlation function, while the latter technique demands that the \ac{TS} provides an ``uncorrectable'' error pattern, ensuring (with high enough probability) that the decoder does not misinterpret it as a codeword. Fortunately, these requirements are not mutually exclusive. 

\subsection{Our contribution}

In this paper, we evaluate the performance of a communication scheme that incorporates error-correcting codes and the \ac{TS}. We assess the \ac{TC} rejection probability both theoretically, where feasible, and numerically using Monte Carlo simulations, otherwise. This metric is crucial for evaluating the system performance as it determines the rate at which \acp{CLTU} are discarded and, consequently, the \ac{TC} cannot be used by the space element.
 In particular, we focus on the \ac{CCSDS}-compliant scenario in which:
\begin{itemize}
    \item data are encoded with an \ac{LDPC} error-correcting code, and consequently decoded by iterative algorithms;
    \item the \ac{TS} is detected by exploiting the trigger of a decoding error.
\end{itemize} 
As for \ac{LDPC} decoding algorithms, we consider the \ac{LLR-SPA} \cite{Hu2001}, the \ac{MSA} \cite{FossoMinsum}, and the \ac{NMSA} \cite{FossoNMS}. We examine the scheme performance by using various numbers of decoding iterations and \ac{CLTU} lengths, showing that, somehow counterintuitively, a small number of decoding iterations can improve performance, from the \ac{TS} detection standpoint, when the \ac{CLTU} is extremely short.

Our study starts from \cite{bluebook} but considers a more general framework.
Moreover, when using the parameters recommended in \cite{bluebook}, our numerical results indicate that the performance of the current system falls short of expectations, seriously mining the reliability of the current standard. To understand this issue, we present a detailed processing of the decoding results. Following such an analysis, we suggest a modification to the standard system aimed at reducing the \ac{TC} rejection probability.
Throughout the paper, we provide many theoretical insights to understand and justify the results obtained.

\subsection{Paper outline} 
The paper is organized as follows. In Section \ref{sec:preli} we establish the necessary context. In Section \ref{sec:tcrej} we analyze the \ac{TC} rejection probability, and provide insights on the single contributions forming it. In Section \ref{sec:numres} we assess the whole system performance and propose some solutions to improve it.
Finally, Section \ref{sec:concl} provides some concluding remarks.

\section{Preliminaries}\label{sec:preli}
  In this section, we introduce the terminology and notation used throughout the paper, and provide a brief overview of the communication system established in \cite{bluebook}.

\subsection{Notation}

The Hamming distance between two vectors is defined as the number of positions at which they differ, and is denoted as $d_\mathrm{H}(\cdot)$. Instead, the Euclidean distance is denoted  as $d_\mathrm{E}(\cdot)$. The Hamming weight of a vector is given by the number of its non-zero entries.

Given the finite field $\mathbb F_2$, an $(n,k)$ linear binary code $\mathcal C$ is a $k-$dimensional subspace of $\mathbb F_2^n$, where $k<n$. The codewords in $\mathcal C$ can be obtained as $ \mathcal C = \{ \mathbf c\in\mathbb F_2^n | \mathbf c \mathbf H^\top = \mathbf 0 \}$,  and $\mathbf H\in\mathbb F_2^{r\times n}$ is a full-rank matrix of size $r \times n$, where $r = n-k$, which is known as the parity-check matrix. The code rate $R$ is defined as $R=\frac{k}{n}$. The number of codewords of Hamming weight $w$ is denoted as $A(w)$, often referred to as weight enumerator function or distance distribution. Since the considered error correcting codes are linear, the minimum Hamming distance of the code, simply denoted as $d_{\min}$, is the smallest positive value of $w$ in the code such that $A(w)>0$. In a linear code, all codewords have identical Hamming distance properties; therefore, $A(w)$ also represents the number of codewords at Hamming distance $w$ from any fixed codeword. \ac{LDPC} codes are characterized by parity-check matrices with a relatively small number of non-zero entries compared to the number of zeros.

\subsection{Standard Communication System}\label{subsec:stdcomsys}
\begin{figure}
    \centering
    \resizebox{0.6\textwidth}{!}{

\begin{tikzpicture}[node distance=2.25cm and 1.75cm, 
    every node/.style={rectangle, draw, text width=4cm, align=center, minimum height=1cm},  auto]

    \node (infoword1) [rectangle, draw=none, fill=none] {Infowords};
    \node (encoder) [below=of infoword1] {Encoder};
    \node (randomizer1) [below=of encoder] {Randomizer};
    \node (cltu) [below=of randomizer1] {Encapsulation of randomized encoded data into a CLTU};
    \node (awgn) [below=of cltu, xshift=3cm] {AWGN}; 

    \node (infoword2) [right=of infoword1, rectangle, draw=none, fill=none] {Decoded infowords};
    \node (decoder) [below=of infoword2] {Decoder};
    \node (randomizer2) [below=of decoder] {De-randomizer};
    \node (detection) [below=of randomizer2] {Detection of start sequence};

    \draw[->] (infoword1) -- (encoder);
    \draw[->] (encoder) -- (randomizer1) node[midway, right, rectangle, draw=none, fill=none,xshift=-0.6cm] {Encoded data};
    \draw[->] (randomizer1) -- (cltu) node[midway, right, rectangle, draw=none, fill=none,xshift=-0.6cm] {Randomized encoded data};
    \draw[->] (cltu) |- (awgn);
    \node[below right=1.09cm and -3.25cm of cltu, rectangle, draw=none, fill=none] {CLTU}; 

    \draw[->] (awgn) -| (detection);
    \node[above right=0.2cm and 0.4cm of awgn, rectangle, draw=none, fill=none] {Noisy CLTU}; 

    \draw[->] (detection) -- (randomizer2) node[midway, right, rectangle, draw=none, fill=none] {Noisy randomized encoded data and noisy tail sequence};
    \draw[->] (randomizer2) -- (decoder) node[midway, right, rectangle, draw=none, fill=none] {Noisy encoded data and randomized noisy tail sequence};
    \draw[->] (decoder) -- (infoword2);

\end{tikzpicture}}
    \caption{Communication System's Blocks Diagram when using \ac{LDPC} codes}
    \label{fig:blockdiag}
\end{figure}
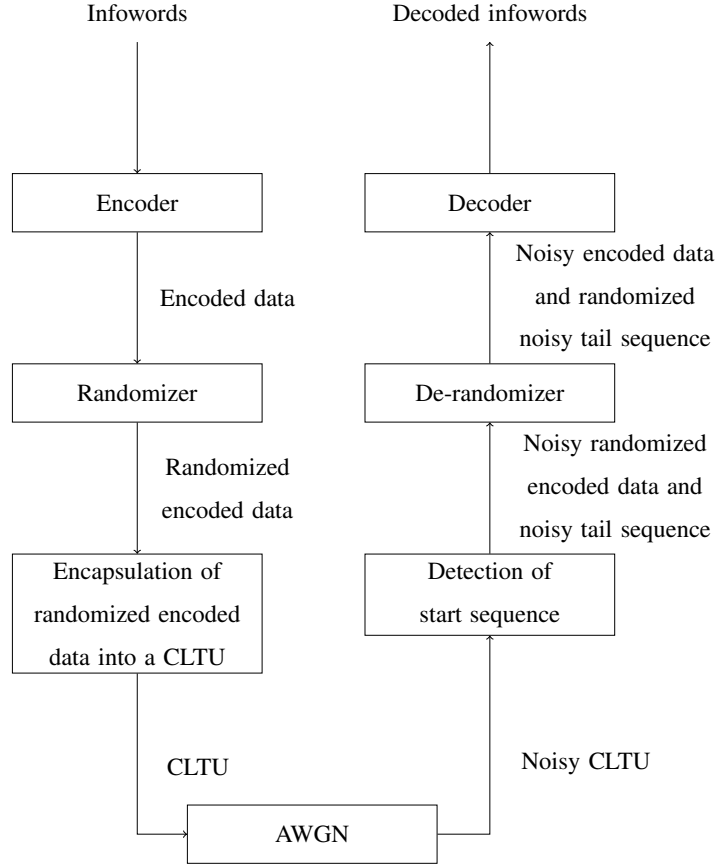

The \ac{TC} communication system described in  \cite{bluebook}, for the case using \ac{LDPC} coding, is summarized in Fig. \ref{fig:blockdiag}. Basically, the information sequences (infowords) are encoded, then the encoded data are randomized and encapsulated into a \ac{CLTU}; in this stage, the start sequence and the \ac{TS} are added, respectively, ahead and behind the encoded data. We remark that, as anticipated, when data are encoded with the $(128,64)$ \ac{LDPC} code (described in depth in Appendix \ref{app:LDPC12864}), the inclusion of the \ac{TS} is optional; instead, for the $(512,256)$ \ac{LDPC} code (its parity-check matrix is shown in Appendix \ref{app:LDPC12864}) the \ac{TS} shall not be used at all (see \cite[Section 5.2.4.3]{bluebook}). For this reason, we do not discuss further the latter code. The $(128,64)$ \ac{LDPC} code code has $R=\frac{1}{2}$ and minimum distance $d_{\mathrm{min}} = 14$ \cite{greenbook}. Note that, if \ac{BCH} coding is used, a different communication scheme should be employed. In particular, in that case, randomization is optional and, if used, it is applied before the encoding operation.

At the receiver side, the start sequence is detected, then the encoded data and the \ac{TS} are input to the de-randomizer (therefore, the encoded data are de-randomized, whereas the \ac{TS} is randomized) and, finally, the decoding process starts. 
The decoding process stops:
\begin{enumerate}
    \item if the \ac{TS} is recognized by a hard/soft correlator (or with any alternative approach specific to the \ac{TS} detection); or
    \item if the decoder reaches a predetermined maximum number of iterations without converging to a codeword. 
\end{enumerate}
In this paper, we focus on the latter approach, assuming that the \ac{TS} is designed as a vector which is sufficiently distant, according to some metric, from the \ac{LDPC} codewords and thus triggers a decoding failure with high probability.

\subsection{Randomized Tail Sequence and De-Randomized Tail Sequence}
\label{subsec:Scenarios}

In the rest of the paper, we will refer to the randomized \ac{TS} case, or simply \emph{randomized case}, as the standard case: at the receiving end, the noisy randomized encoded data, along with the non-randomized \ac{TS}, are both given as the input to the  de-randomizer, so that, as observed above, the encoded data get de-randomized, and the \ac{TS} is randomized, since it was appended to the encoded data at the transmitter side without undergoing randomization. As demonstrated in the following sections, this approach introduces an anomaly that significantly undermines the system's performance.

On the other hand, for the aforementioned reasons, in this paper we also propose an alternative solution, out of the standard, where, at the transmitting side, the \ac{TS} is also randomized along with the encoded data. We call this option the de-randomized \ac{TS} case or, simply, the ``\textit{de-randomized case}". In fact, this way, at the receiving side, the \ac{TS} gets de-randomized as well.

\section{TeleCommand Rejection Probability} \label{sec:tcrej}
 To analyze the system performance, by generalizing \cite[Equation (14)]{tutorial}, we compute the \ac{TC} rejection probability, which is the probability that the \ac{TC} gets rejected from the satellite, when \ac{LDPC} coding is employed, and the \ac{TS} ends the \ac{CLTU}.

  The \ac{TC} rejection probability of the communication system described in Fig. \ref{fig:blockdiag} can be computed as 
   \begin{equation}
   P_{\mathrm{TCrej}} =P_{\mathrm{nat}}+(1 - P_{\mathrm{nat}}) [P_{\mathrm{md}}+(1-P_{\mathrm{md}}) P_{\mathrm{LDPC}}] ,
   \label{eq:P_TCrej}
\end{equation}   
where
\begin{itemize}
    \item $P_{\mathrm{md}}$ is the \emph{missed detection probability}, i.e., the probability that the start sequence is not detected;

    \item $P_{\mathrm{nat}}$ is the \emph{not-acknowledged termination probability}, that is, the probability that the termination of the \ac{CLTU} (which is the \ac{TS}) is not recognized;

    \item $P_{\mathrm{LDPC}}$ is the probability that decoding fails for any of the \ac{LDPC} codewords, and the error is detected.
\end{itemize}

    In fact, following the order given in Fig. \ref{fig:CLTU}, we readily notice that the \ac{TC} is rejected if
    \begin{itemize}
        \item the start sequence is not correctly recognized, or if
        \item the start sequence is correctly detected, but the decoder fails (in a detectable way) to decode any of the $N$ codewords, or if
        \item the start sequence is correctly detected, and the decoder decodes all the $N$ codewords, but the \ac{TS} is not recognized.
    \end{itemize}
    Since these events are mutually exclusive, the corresponding probabilities can be summed, leading to 
    \[
   P_{\mathrm{TCrej}} = P_{\mathrm{md}} + (1 - P_{\mathrm{md}}) [P_{\mathrm{LDPC}}+(1-P_{\mathrm{LDPC}}) P_{\mathrm{nat}} ].
\]  
Equation \eqref{eq:P_TCrej} is obtained by solving further the above equation. It is easy to verify that the dominant contribution in \eqref{eq:P_TCrej} is  $P_{\mathrm{md}} + P_{\mathrm{nat}} + P_{\mathrm{LDPC}}$, since the other terms are products of two or three probabilities. So, in the following, we will discuss the behavior of these leading components, reasoning on their impact on the overall probability $P_{\mathrm{TCrej}}$. 

It is obvious to observe that, when $P_{\mathrm{md}}$ and $P_{\mathrm{LDPC}}$ tend to $0$ (which occurs for high values of the signal-to-noise ratio), $P_{\mathrm{TCrej}}$ converges to $P_{\mathrm{nat}}$. This straightforward reasoning will help justifying the performance results, presented in Section \ref{sec:numres}.

Let us study separately the three leading probabilities contributing to $P_{\mathrm{TCrej}}$, and provide more thorough definitions.

\subsection{Missed detection probability}

The missed detection probability $P_{\mathrm{md}}$ is the probability that the start sequence is not correctly detected, and therefore the receiver does not recognize the beginning of the \ac{CLTU}. In order to model $P_{\mathrm{md}}$, we consider that an $S$-bits start sequence, along with the \ac{CLTU}, is \ac{BPSK}-modulated and transmitted over the \ac{AWGN} channel. Moreover, for the sake of ease, we assume that a bit-by-bit comparison of the (hard) received sequence with the actual start sequence is employed for detection, and that the received sequence is accepted as the start sequence if it differs in up to $E$ positions from the actual start sequence. In other words, the metric we consider is the number of positions in which the tentative sequence and the start sequence match. To be noticed that, if the start sequence is well-designed, this approach corresponds to the use of a hard correlator. The missed detection probability can then be computed, by extending the formulation in \cite{tutorial}, as 
  \begin{equation}
   P_{\mathrm{md}} = 1-\sum_{j=0}^E {S \choose j} P_b^j (1-P_b)^{S-j},
    \label{eq:eqpmd}
\end{equation} 
where $P_b$ is the bit error probability for \ac{BPSK}-modulated transmissions over the \ac{AWGN} channel, that is, $P_b=\frac{1}{2}\mathrm{erfc}\left(\sqrt{\frac{E_b}{N_0}}\right)$,
being $\mathrm{erfc}(\cdot)$ the complementary error function and $\frac{E_b}{N_0}$ the signal-to-noise ratio per bit. Equation \eqref{eq:eqpmd} is obtained by considering that the start sequence is correctly detected if at most $E$  hard errors occurred during transmission.

\subsection{Probability of decoding failure on coded data} \label{subsec:LDPCdecodefail}

The \ac{TC} gets rejected also if the decoder fails to converge to a codeword while decoding the \ac{LDPC}-coded data. Also in this case, we actually need to consider two possible scenarios, corresponding to different  decoding errors:
\begin{itemize}
    \item the decoder does not produce a codeword as output, resulting in a \emph{detectable error};
    \item the decoder converges to a codeword that is not the transmitted one, yielding an \emph{undetectable error}.
\end{itemize}
If an undetectable error occurs, the \ac{TC} would not be rejected. Therefore, the undetectable error rate should not  be considered in the computation of $P_{\mathrm{TCrej}}$.

Denoting the \ac{CER} as $\mathrm{CER}$ the probability that a codeword is incorrectly decoded, we can write 
\[
\mathrm{CER}=\mathrm{CER}^*+\mathrm{UCER},
\]
being $\mathrm{CER}^*$ the ``detectable'' \ac{CER} and $\mathrm{UCER}$ the ``undetectable'' \ac{CER}.
Given this, assuming that the \ac{CLTU} contains $N$ codewords, based on its definition, we can compute 
\begin{equation}
P_{\mathrm{LDPC}}=1-(1-\mathrm{CER}^*)^N.
\label{eq:PLDPC}
\end{equation}
For practical error-correcting codes and not extremely low values of $E_b/N_0$ (see, for example, \cite[Fig. 11-9]{greenbook}) it is possible to assume
\[
\mathrm{CER}^*\approx \mathrm{CER},
\]
since $\mathrm{UCER}\ll  \mathrm{CER}^*$ and is thus negligible. The value of the \ac{CER} will be estimated through Monte Carlo simulations as\footnote{Here and in the following, $\#$ conventionally reads as ``Number of''.}
\begin{equation}
    \mathrm{CER}\approx \frac{\#_{\mathrm{decoding-failures}}}{\#_{\mathrm{decoding-attempts}}}.
    \label{eq:PLDPCmc}
\end{equation}

It is interesting to study the behavior of  \eqref{eq:PLDPC} when $N$ is small to moderate and $E_b/N_0$ is large. In this setting, it is possible to approximate $P_{\mathrm{LDPC}}$ as $1 - (1 - \mathrm{CER})^N$. In particular, if the \ac{CER} is small and \(N\) is not too large, we can write
   \begin{equation}
      P_{\mathrm{LDPC}} \approx 1 - (1 - \mathrm{CER})^N \approx N \cdot \mathrm{CER}. 
   \label{eq:approximationCER}
   \end{equation}

\subsection{Not-Acknowledged termination probability}

Assuming that the \ac{TS} is detected by triggering a detectable decoding error, the reasoning on the \ac{TC} rejection probability is opposite to that of an \ac{LDPC} decoding failure. In this case, when the noisy (randomized) \ac{TS} is fed to the decoder, we expect it to fail. If the decoder mistakenly converges to a codeword, it would indicate that it has incorrectly identified the \ac{TS} as valid encoded data, which is clearly undesirable. 
We can thus estimate $P_{\mathrm{nat}}$ through Monte Carlo simulations, where the \ac{TS} is the object of the transmission; however, somehow counterintuitively, we have a not-acknowledged termination when the decoder succeeds, whereas the termination is correctly acknowledged if the decoder fails. In other words, the Probability of Not-Acknowledged Termination is estimated as
\begin{equation}
 P_{\mathrm{nat}} =  \frac{\#_{\mathrm{decoding-successes}}}{\#_{\mathrm{decoding-attempts}}}=1-\frac{\#_{\mathrm{decoding-failures}}}{\#_{\mathrm{decoding-attempts}}}.   
 \label{eq:pnat}
\end{equation}
We thus notice that the not-acknowledged termination probability is the complementary of the codeword error rate  \eqref{eq:PLDPCmc}.

In order to minimize the likelihood of a decoding success, the \ac{TS} can be designed as a pattern that is as far as possible from any codeword of the considered \ac{LDPC} code, according to the \ac{ML} decoding principle for the \ac{AWGN} channel. The latter states that, keeping in mind the mapping $\mathbf{x}=(-1)^\mathbf{c}$,  and receiving $\mathbf{y}$ as input (which is the noisy \ac{TS}, in our setting), the decoder output is the following estimated codeword $\hat{\mathbf{c}}$:
\[
\hat{\mathbf{c}}=\mathrm{arg}\,\min_{\mathbf{c}}\, d_\mathrm{E}(\mathbf{y,x}). 
\]
Clearly, since the \ac{TS} is a binary vector, a good strategy consists in designing the \ac{TS} in such a way that its Hamming distance from all the  $2^{k}$ codewords (denoted as $\mathbf{c}_i$, with $0\leq i \leq 2^k-1$) of the $(128,64)$ \ac{LDPC} code is the largest possible, i.e., if we denote the \ac{TS} as $\mathbf{t}$, in such a way that
$\min_{0\leq i \leq 2^k-1} d_\mathrm{H}(\mathbf{t},\mathbf{c}_i)$ is maximized. It is important to remark that the practical iterative decoders commonly used for decoding \ac{LDPC} codes are suboptimal compared to \ac{ML} and, most importantly, they are not complete decoders. As a result, when iterative decoders (such as those based on the \ac{LLR-SPA}, the \ac{MSA}, and the \ac{NMSA}, considered in Section \ref{sec:numres}-\ref{subsec:randotc}) are given an input that is significantly distant from any codeword, they are expected to be unable to converge to a codeword and thus return a decoding failure.

\section{Performance Analysis and Improvement}\label{sec:numres}

In this section we delve into the analysis of the performance of the communication system in Fig. \ref{fig:blockdiag}, considering the parameters assumed in \cite{bluebook}, and then propose a solution to improve its  performance in terms of \ac{TC} rejection probability. 

\subsection{TC rejection probability in the randomized case\label{subsec:randotc}}

According to Section \ref{sec:tcrej}, the \ac{TC} rejection probability results from three contributions. Following the \ac{CCSDS} recommendations, when \ac{LDPC} coding is employed, the start sequence consists of the following $64$-bit pattern: $\mathrm{0347\,76C7\,2728\,95B0}$ (in hexadecimal); the \ac{TS} is instead formed by the following $128$-bit pattern (in hexadecimal): 
\[\mathbf{t} = \mathrm{5555\,5556\,AAAA\,AAAA\,5555\,5555\,5555\,5555}.\] For the sake of results' reproducibility, we remind that the considered randomizer exploits a \ac{LFSR} characterized by the polynomial $x^8+x^6+x^4+x^3+x^2+x+1$, which generates a pseudo-random sequence of period $255$ that is summed modulo $2$ to the input. When de-randomizing, the same operations of the randomization phase are applied. The standard recommends resetting the \ac{LFSR} before randomizing or de-randomizing each input $128$-bit sequence. Consequently, the exact randomizing (or de-randomizing) sequence XOR-ed with the codewords and the \ac{TS} is always the same and is known; it is determined by the first $128$ bits generated by the above \ac{LFSR}.

At first, we assume that the \ac{CLTU} only contains one codeword, i.e., $N=1$. This assumption is extreme and optimistic, since $P_{\mathrm{LDPC}}$ gets larger for increasing values of $N$, according to \eqref{eq:PLDPC}. In fact, for any integer $N>1$
\begin{equation}
   \mathrm{CER} < 1-(1-\mathrm{CER})^N, 
   \label{eq:cerN}
\end{equation}
being $0\leq \mathrm{CER}\leq 1$.
Therefore, the curves we obtain in the rest of this section  represent a lower bound on the actual \ac{TC} rejection probability. Later, in Section \ref{sec:numres}-\ref{subsec:manycods}, we will consider larger and more practical values of $N$.

As mentioned in the Introduction, we consider three decoding algorithms that are commonly used for decoding of \ac{CCSDS}-compliant \ac{LDPC} codes: the \ac{LLR-SPA}, the \ac{MSA}, and the \ac{NMSA} with normalization factor equal to $0.8$. For all these algorithms, we consider the scenarios in which the decoder runs at most $N_{\rm{it}} = 100$ or $N_{\rm{it}} = 20$ decoding iterations. We have performed Monte Carlo simulations for both the case of noisy codewords and noisy (randomized or de-randomized) \ac{TS} at the input of the decoder, stopping the simulations upon encountering $100$ decoding errors (when decoding encoded data) and $100$ decoding successes (when decoding the \ac{TS}).

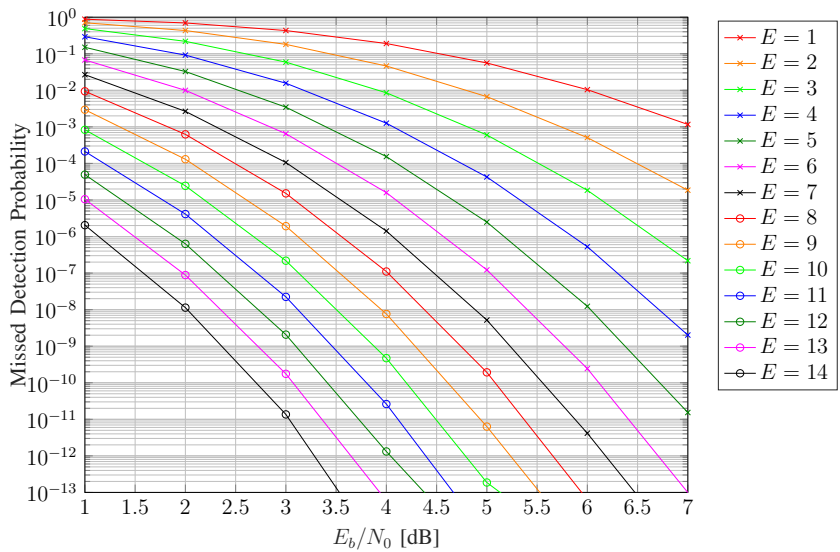
\begin{figure}[t!]
    \centering
    \resizebox{0.7\textwidth}{!}{
    \definecolor{mycolor1}{rgb}{1.00000,1.00000,0.00000}%
\definecolor{mycolor2}{rgb}{0.00000,1.00000,1.00000}%
\definecolor{mycolor3}{rgb}{1.00000,0.00000,1.00000}%
\definecolor{darkgreen}{rgb}{0.0, 0.5, 0.0}

\begin{tikzpicture}

\begin{axis}[%
width=4.521in,
height=3.566in,
at={(0.758in,0.481in)},
scale only axis,
xmin=1,
xmax=7,
xlabel style={font=\color{white!15!black}},
xlabel={$E_b/N_0$ [dB]},
ymode=log,
ymin=1e-13,
ymax=1,
yminorticks=true,
ylabel style={font=\color{white!15!black}},
ylabel={Missed Detection Probability},
axis background/.style={fill=white},
xmajorgrids,
ymajorgrids,
yminorgrids,
legend style={at={(1.05,0.215)}, anchor=south west, legend cell align=left, align=left, draw=white!15!black, fill opacity=0.7}
]
]
\addplot [color=red, mark=x, mark options={solid, red}]
  table[row sep=crcr]{%
0	0.965830360038001\\
1	0.881785918101847\\
2	0.697458850168961\\
3	0.431944634149008\\
4	0.190767200417948\\
5	0.0560602014754372\\
6	0.0104236941902831\\
7	0.0011658364536445\\
};
\addlegendentry{$E=1$}

\addplot [color=orange, mark=x, mark options={solid, orange}]
  table[row sep=crcr]{%
0	0.888165800935449\\
1	0.705811719424091\\
2	0.432383512854315\\
3	0.180665770561645\\
4	0.046341142932104\\
5	0.00670939046555385\\
6	0.000508978080280165\\
7	1.85529249665459e-05\\
};
\addlegendentry{$E=2$}

\addplot [color=green, mark=x, mark options={solid, green}]
  table[row sep=crcr]{%
0	0.751151775329132\\
1	0.488918327013865\\
2	0.218909988509423\\
3	0.0590741698091044\\
4	0.00855629487914988\\
5	0.000600570466833594\\
6	1.84358711687038e-05\\
7	2.1823171085511e-07\\
};
\addlegendentry{$E=3$}

\addplot [color=blue, mark=x, mark options={solid, blue}]
  table[row sep=crcr]{%
0	0.572788030985617\\
1	0.291656497912753\\
2	0.09205191072985\\
3	0.0156580905328924\\
4	0.00126190095736101\\
5	4.25890031210963e-05\\
6	5.26848320442364e-07\\
7	2.02162594303834e-09\\
};
\addlegendentry{$E=4$}

\addplot [color=darkgreen, mark=x, mark options={solid, darkgreen}]
  table[row sep=crcr]{%
0	0.39007920060512\\
1	0.150483644658207\\
2	0.0327315702424262\\
3	0.00345951753147544\\
4	0.000153818324387101\\
5	2.48445590456114e-06\\
6	1.2356113465728e-08\\
7	1.53546966807916e-11\\
};
\addlegendentry{$E=5$}

\addplot [color=mycolor3, mark=x, mark options={solid, mycolor3}]
  table[row sep=crcr]{%
0	0.236712169927107\\
1	0.0676934613560395\\
2	0.0100010875107008\\
3	0.000650936407415204\\
4	1.58832864487479e-05\\
5	1.22417370462813e-07\\
6	2.44410075511681e-10\\
7	9.93233273405281e-14\\
};
\addlegendentry{$E=6$}

\addplot [color=black, mark=x, mark options={solid, black}]
  table[row sep=crcr]{%
0	0.128236167162549\\
1	0.0267828764975568\\
2	0.00266198632920656\\
3	0.000106064887539614\\
4	1.41536169384615e-06\\
5	5.19526122122471e-09\\
6	4.16147671877809e-12\\
7	1.58206781009085e-15\\
};
\addlegendentry{$E=7$}

\addplot [color=red, mark=o, mark options={solid, red}]
  table[row sep=crcr]{%
0	0.0622595182950515\\
1	0.00939896719618916\\
2	0.000624324985772806\\
3	1.51665042925631e-05\\
4	1.10414916210111e-07\\
5	1.9276236162824e-10\\
6	6.34770014329433e-14\\
7	1.04083408558608e-15\\
};
\addlegendentry{$E=8$}

\addplot [color=orange, mark=o, mark options={solid, orange}]
  table[row sep=crcr]{%
0	0.0272160178308066\\
1	0.00294807394137231\\
2	0.000130262605113618\\
3	1.92373362351006e-06\\
4	7.62748653126977e-09\\
5	6.32838226266585e-12\\
6	2.44249065417534e-15\\
7	1.04083408558608e-15\\
};
\addlegendentry{$E=9$}

\addplot [color=green, mark=o, mark options={solid, green}]
  table[row sep=crcr]{%
0	0.0107631396912282\\
1	0.000832104284719026\\
2	2.4374119100723e-05\\
3	2.18363188309034e-07\\
4	4.70925409778999e-10\\
5	1.86795023893183e-13\\
6	1.63757896132211e-15\\
7	1.04083408558608e-15\\
};
\addlegendentry{$E=10$}

\addplot [color=blue, mark=o, mark options={solid, blue}]
  table[row sep=crcr]{%
0	0.00386845092651145\\
1	0.000212609862066504\\
2	4.1181010136393e-06\\
3	2.2344722294676e-08\\
4	2.61843879911794e-11\\
5	6.21724893790088e-15\\
6	1.63757896132211e-15\\
7	1.04083408558608e-15\\
};
\addlegendentry{$E=11$}

\addplot [color=darkgreen, mark=o, mark options={solid, darkgreen}]
  table[row sep=crcr]{%
0	0.00126900372632976\\
1	4.94328731713845e-05\\
2	6.31895685443418e-07\\
3	2.07402017782243e-09\\
4	1.31850086404484e-12\\
5	1.4432899320127e-15\\
6	1.63757896132211e-15\\
7	1.04083408558608e-15\\
};
\addlegendentry{$E=12$}

\addplot [color=mycolor3, mark=o, mark options={solid, mycolor3}]
  table[row sep=crcr]{%
0	0.000381413109508899\\
1	1.05063337376698e-05\\
2	8.84987106930524e-08\\
3	1.75540471047952e-10\\
4	5.93969318174459e-14\\
5	1.33226762955019e-15\\
6	1.63757896132211e-15\\
7	1.04083408558608e-15\\
};
\addlegendentry{$E=13$}

\addplot [color=black, mark=o, mark options={solid, black}]
  table[row sep=crcr]{%
0	0.000105402005977195\\
1	2.0493630045193e-06\\
2	1.13615552521651e-08\\
3	1.3611223259602e-11\\
4	1.33226762955019e-15\\
5	1.33226762955019e-15\\
6	1.63757896132211e-15\\
7	1.04083408558608e-15\\
};
\addlegendentry{$E=14$}

\end{axis}

\begin{axis}[%
width=5.833in,
height=4.375in,
at={(0in,0in)},
scale only axis,
xmin=0,
xmax=1,
ymin=0,
ymax=1,
axis line style={draw=none},
ticks=none,
axis x line*=bottom,
axis y line*=left
]
\end{axis}
\end{tikzpicture}%
}
    \caption{Missed detection probability under hard correlation detection}
    \label{fig:pmd}
\end{figure}

Let us start from the missed detection probability. In Fig. \ref{fig:pmd} we show the behavior of \eqref{eq:eqpmd}, as a function of the signal-to-noise ratio per bit, computed for $S = 64$ and many different values of $E$. If ``soft" detection approaches were employed, we could expect even lower missed detection probabilities (see \cite[Fig. 5]{Baldi2016}). As we will demonstrate in the following, when $E$ and $E_b/N_0$ are sufficiently large, say greater than $8$ and $1.5$ dB, respectively, the impact of $P_{\mathrm{md}}$ on $P_{\mathrm{TCrej}}$ becomes negligible, as it is much smaller than $P_{\mathrm{LDPC}}$ and $P_{\mathrm{nat}}$. For reference, we will use $E=13$, yielding the same results as in \cite[Fig. 5]{Baldi2016}, for the hard-correlated case.

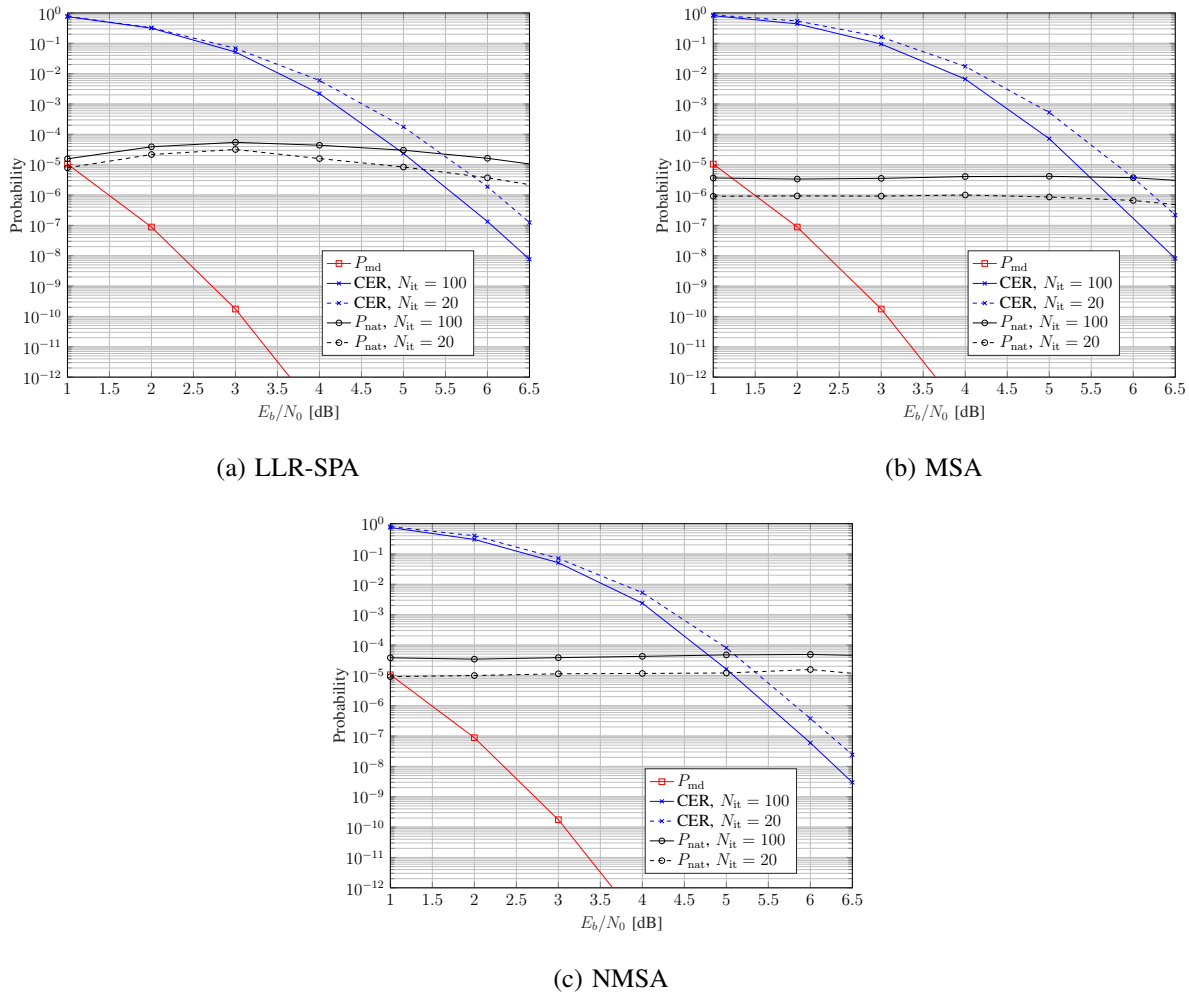
\begin{figure*}
    \centering
    \begin{subfigure}[b]{0.48\textwidth}
        \centering
        \resizebox{\textwidth}{!}{\begin{tikzpicture}

\begin{axis}[%
width=4.521in,
height=3.566in,
at={(0.758in,0.481in)},
scale only axis,
unbounded coords=jump,
xmin=1,
xmax=6.5,
xlabel style={font=\color{white!15!black}},
xlabel={$E_b/N_0$ [dB]},
ymode=log,
ymin=1e-12,
ymax=1,
yminorticks=true,
ylabel style={font=\color{white!15!black}},
ylabel={Probability},
axis background/.style={fill=white},
xmajorgrids,
ymajorgrids,
yminorgrids,
legend style={at={(0.551,0.059)}, anchor=south west, legend cell align=left, align=left, draw=white!15!black}
]
\addplot [color=red, mark=square, mark options={solid, red}]
  table[row sep=crcr]{%
0   0.000381413109508899\\
1	1.05063337376698e-05\\
2	8.84987106930524e-08\\
3	1.75540471047952e-10\\
4	5.93969318174459e-14\\
5	1.33226762955019e-15\\
6	1.63757896132211e-15\\
7	1.04083408558608e-15\\
};
\addlegendentry{$P_{\mathrm{md}}$}

\addplot [color=blue, mark=x, mark options={solid, blue}]
  table[row sep=crcr]{%
1 7.576E-01\\
2 3.155E-01\\
3 5.173E-02\\
4 2.203E-03\\
5 2.358E-05\\
6 1.34E-07\\
6.5 7.69E-09\\
};
\addlegendentry{CER, $N_{\mathrm{it}}=100$}

\addplot [color=blue, dashed, mark=x, mark options={solid, blue}]
  table[row sep=crcr]{%
1 7.75E-01\\
2 3.268E-01\\
3 6.849E-02\\
4 5.986E-03\\
5 1.773E-04\\
6 1.89E-06\\
6.5 1.246E-07\\
7 8.336E-09\\
};
\addlegendentry{CER, $N_{\mathrm{it}}=20$}

\addplot [color=black, mark=o, mark options={solid, black}]
  table[row sep=crcr]{%
0	2.00000466667756e-06\\ 
1	1.563E-05\\ 
2	3.902E-05\\ 
3	5.458e-05\\
4	4.383e-05\\
5	3.010e-05\\
6	1.628e-05\\
7	6.905e-06\\ 
8   3.321E-06\\
};
\addlegendentry{$P_{\mathrm{nat}}$, $N_{\mathrm{it}}=100$}

\addplot [color=black, dashed, mark=o, mark options={solid, black}]
  table[row sep=crcr]{%
1	7.91E-06\\ 
2	2.168E-05\\ 
3	3.175E-05\\ 
4	1.591E-05\\ 
5	8.39E-06 \\ 
6 3.713E-06 \\ 
7 1.306E-06 \\ 
};
\addlegendentry{$P_{\mathrm{nat}}$, $N_{\mathrm{it}}=20$}

\end{axis}

\begin{axis}[%
width=5.833in,
height=4.375in,
at={(0in,0in)},
scale only axis,
xmin=0,
xmax=1,
ymin=0,
ymax=1,
axis line style={draw=none},
ticks=none,
axis x line*=bottom,
axis y line*=left
]
\end{axis}
\end{tikzpicture}%
}
        \caption{\ac{LLR-SPA}}
        \label{fig:component_comp}
    \end{subfigure}
    \hfill
    \begin{subfigure}[b]{0.48\textwidth}
        \centering
        \resizebox{\textwidth}{!}{\begin{tikzpicture}

\begin{axis}[%
width=4.521in,
height=3.566in,
at={(0.758in,0.481in)},
scale only axis,
unbounded coords=jump,
xmin=1,
xmax=6.5,
xlabel style={font=\color{white!15!black}},
xlabel={$E_b/N_0$ [dB]},
ymode=log,
ymin=1e-12,
ymax=1,
yminorticks=true,
ylabel style={font=\color{white!15!black}},
ylabel={Probability},
axis background/.style={fill=white},
xmajorgrids,
ymajorgrids,
yminorgrids,
legend style={at={(0.551,0.059)}, anchor=south west, legend cell align=left, align=left, draw=white!15!black}
]
\addplot [color=red, mark=square, mark options={solid, red}]
  table[row sep=crcr]{%
0   0.000381413109508899\\
1	1.05063337376698e-05\\
2	8.84987106930524e-08\\
3	1.75540471047952e-10\\
4	5.93969318174459e-14\\
5	1.33226762955019e-15\\
6	1.63757896132211e-15\\
7	1.04083408558608e-15\\
};
\addlegendentry{$P_{\mathrm{md}}$}

\addplot [color=blue, mark=x, mark options={solid, blue}]
  table[row sep=crcr]{%
1 8.13E-01\\
2 4.384E-01\\
3 9.477E-02\\
4 6.633E-03\\
5 7.216E-05\\
6.5 8.22E-09  \\
};
\addlegendentry{CER, $N_{\mathrm{it}}=100$}

\addplot [color=blue, dashed, mark=x, mark options={solid, blue}]
  table[row sep=crcr]{%
1 8.62E-01 \\
2 5.45E-01\\
3 1.61E-01\\
4 1.739E-02\\
5 5.284E-04 \\
6 3.7E-06\\
6.5 2.18E-07\\
7 1.052E-08\\
};
\addlegendentry{CER, $N_{\mathrm{it}}=20$}

\addplot [color=black, mark=o, mark options={solid, black}]
  table[row sep=crcr]{%
0	5.33336355572682e-06\\ 
1	3.65617077534951e-06\\
2	3.367E-06\\ 
3	3.55065357057806e-06\\
4	4.07575724105145e-06\\
5	4.147e-06\\
6	3.73572760598474e-06\\
7	2.51071649119939e-06\\
8 1.529E-06\\
};
\addlegendentry{$P_{\mathrm{nat}}$, $N_{\mathrm{it}}=100$}

\addplot [color=black, dashed, mark=o, mark options={solid, black}]
  table[row sep=crcr]{%
1	9.143E-07 \\ 
2 9.38E-07\\ 
3 9.242E-07\\
4 1.003E-06\\ 
5 8.639E-07 \\ 
6 6.646E-07 \\ 
7 3.536E-07\\ 
};
\addlegendentry{$P_{\mathrm{nat}}$, $N_{\mathrm{it}}=20$}

\end{axis}

\begin{axis}[%
width=5.833in,
height=4.375in,
at={(0in,0in)},
scale only axis,
xmin=0,
xmax=1,
ymin=0,
ymax=1,
axis line style={draw=none},
ticks=none,
axis x line*=bottom,
axis y line*=left
]
\end{axis}
\end{tikzpicture}%
}
        \caption{\ac{MSA}}
        \label{fig:component_comp_rand_ms}
    \end{subfigure}
    
    \begin{subfigure}[b]{0.48\textwidth}
        \centering
        \resizebox{\textwidth}{!}{\begin{tikzpicture}

\begin{axis}[%
width=4.521in,
height=3.566in,
at={(0.758in,0.481in)},
scale only axis,
unbounded coords=jump,
xmin=1,
xmax=6.5,
xlabel style={font=\color{white!15!black}},
xlabel={$E_b/N_0$ [dB]},
ymode=log,
ymin=1e-12,
ymax=1,
yminorticks=true,
ylabel style={font=\color{white!15!black}},
ylabel={Probability},
axis background/.style={fill=white},
xmajorgrids,
ymajorgrids,
yminorgrids,
legend style={at={(0.551,0.039)}, anchor=south west, legend cell align=left, align=left, draw=white!15!black}
]
\addplot [color=red, mark=square, mark options={solid, red}]
  table[row sep=crcr]{%
0   0.000381413109508899\\
1	1.05063337376698e-05\\
2	8.84987106930524e-08\\
3	1.75540471047952e-10\\
4	5.93969318174459e-14\\
5	1.33226762955019e-15\\
6	1.63757896132211e-15\\
7	1.04083408558608e-15\\
};
\addlegendentry{$P_{\mathrm{md}}$}

\addplot [color=blue, mark=x, mark options={solid, blue}]
  table[row sep=crcr]{%
1 7.46E-01 \\
2 3.02E-01\\
3 5.17E-02\\
4 2.4E-03\\
5 1.6E-05 \\
6 5.99E-08\\
6.5 2.973E-09\\
};
\addlegendentry{CER, $N_{\mathrm{it}}=100$}

\addplot [color=blue, dashed,  mark=x, mark options={solid, blue}]
  table[row sep=crcr]{%
1 8E-01 \\
2 4E-01\\
3 7.184E-02\\
4 5.33E-03\\
5 8.05E-05 \\
6 3.8E-07\\
6.5 2.436E-08\\
};
\addlegendentry{CER, $N_{\mathrm{it}}=20$}

\addplot [color=black, mark=o, mark options={solid, black}]
  table[row sep=crcr]{%
1	3.81E-05 \\ 
2	3.43E-05 \\ 
3	3.847E-05 \\
4	4.24E-05\\ 
5	4.733E-05\\ 
6	4.89E-05\\ 
7	4.303E-05\\ 
8 2.685E-05\\ 
};
\addlegendentry{$P_{\mathrm{nat}}$, $N_{\mathrm{it}}=100$}

\addplot [color=black, dashed, mark=o, mark options={solid, black}]
  table[row sep=crcr]{%
0	3.93348936174468e-05\\ 
1	9.1e-06\\
2	9.9e-06\\
3	1.13e-05\\
4	1.16e-05\\
5	1.2E-05\\
6	1.56E-05\\
7	8.78E-06\\
};
\addlegendentry{$P_{\mathrm{nat}}$, $N_{\mathrm{it}}=20$}

\end{axis}

\begin{axis}[%
width=5.833in,
height=4.375in,
at={(0in,0in)},
scale only axis,
xmin=0,
xmax=1,
ymin=0,
ymax=1,
axis line style={draw=none},
ticks=none,
axis x line*=bottom,
axis y line*=left
]
\end{axis}
\end{tikzpicture}%
}
        \caption{\ac{NMSA}}
        \label{fig:component_comp_rand_nms}
    \end{subfigure}
    
    \caption{Comparison of \ac{TC} rejection probability's leading components for different algorithms (randomized case)}
    \label{fig:combined}
\end{figure*}

The results of the Monte Carlo simulations for $P_{\mathrm{LDPC}}$ and $P_{\mathrm{nat}}$  are shown in Fig.s \ref{fig:component_comp}, \ref{fig:component_comp_rand_ms}, and \ref{fig:component_comp_rand_nms}, for the randomized case. For better readability, the results are grouped on the basis of the decoder. As anticipated, it is notable that \(P_{\mathrm{md}}\) is always much smaller than the other leading components and does not play a significant role. Additionally, the probability of not-acknowledged termination varies very slowly with increasing values of \(E_b/N_0\).

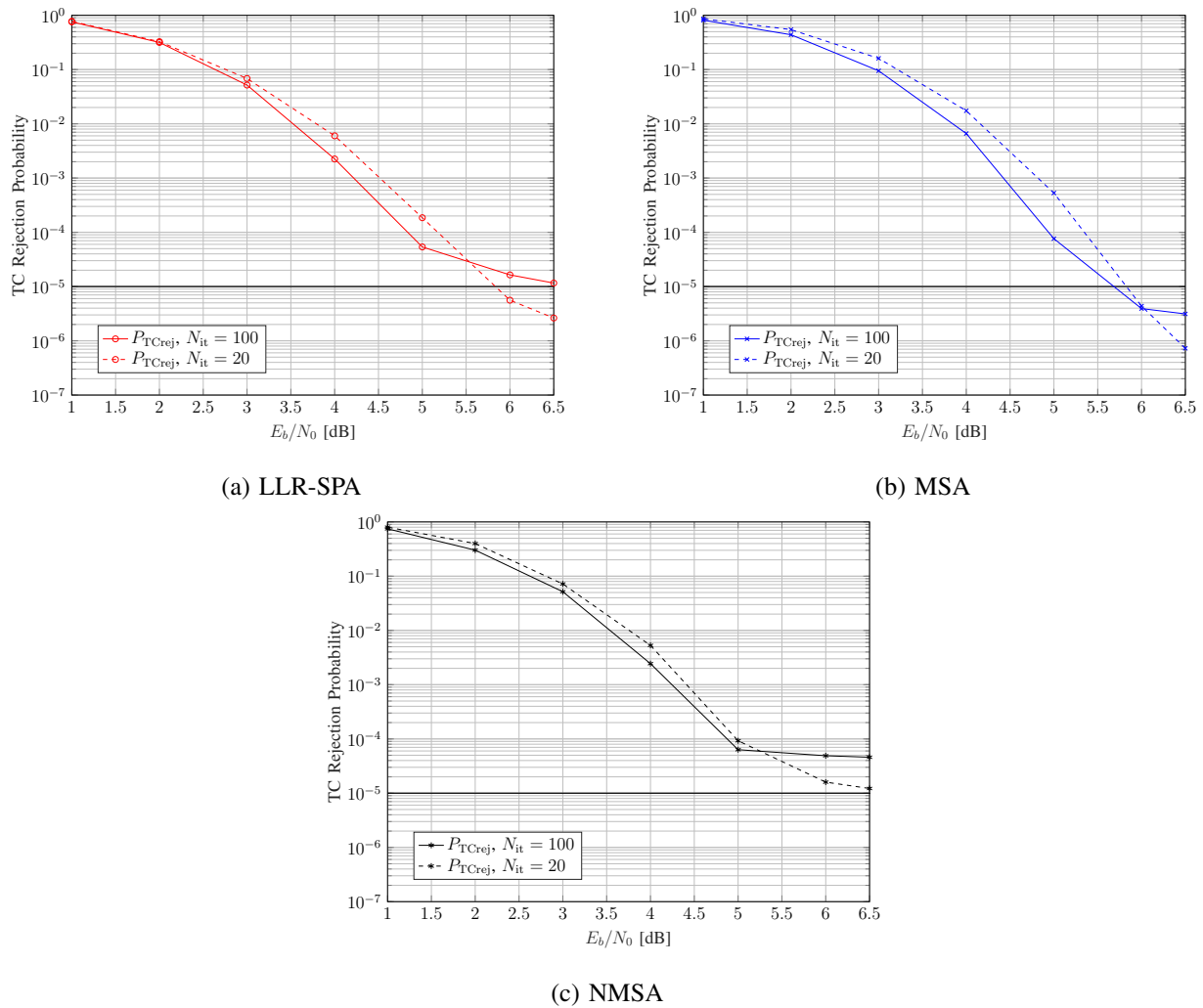
\begin{figure*}
    \centering
    \begin{subfigure}[b]{0.48\textwidth}
        \centering
        \resizebox{\textwidth}{!}{
        \begin{tikzpicture}
        \begin{axis}[%
        width=4.521in,
        height=3.566in,
        scale only axis,
        unbounded coords=jump,
        xmin=1,
        xmax=6.5,
        xlabel style={font=\color{white!15!black}},
        xlabel={$E_b/N_0$ [dB]},
        ymode=log,
        ymin=1e-07,
        ymax=1,
        yminorticks=true,
        ylabel style={font=\color{white!15!black}},
        ylabel={TC Rejection Probability},
        axis background/.style={fill=white},
        title style={font=\bfseries},
        xmajorgrids,
        ymajorgrids,
        yminorgrids,
        legend style={at={(0.054,0.058)}, anchor=south west, legend cell align=left, align=left, draw=white!15!black}
        ]

    \draw[black, thick] (axis cs:1,1e-5) -- (axis cs:7,1e-5);
        
        \addplot [color=red, mark=o, mark options={solid, red}]
          table[row sep=crcr]{%
        0	1\\
        1	0.758006583890304\\
        2	0.31602817293063\\
        3	0.051775099964566\\
        4	0.002247768005941\\
        5	5.3699242441289e-05\\
        6	1.6414e-05\\
        6.5	1.16e-05\\
        };
        \addlegendentry{$P_{\mathrm{TCrej}}$, $N_{\mathrm{it}}=100$}

        \addplot [color=red, dashed, mark=o, mark options={solid, red}]
          table[row sep=crcr]{%
        0	1\\
        1	0.775004208905707\\
        2	0.327018567057995\\
        3	0.0685191890635125\\
        4	0.00600779277905918\\
        5	0.00018569\\
        6	5.603e-06\\
        6.5	2.6341e-06\\
        };
        \addlegendentry{$P_{\mathrm{TCrej}}$,  $N_{\mathrm{it}}=20$}
        \end{axis}
        \end{tikzpicture}
        }
        \caption{\ac{LLR-SPA}}
    \end{subfigure}
    \hfill
    \begin{subfigure}[b]{0.48\textwidth}
        \centering
        \resizebox{\textwidth}{!}{
        \begin{tikzpicture}
        \begin{axis}[%
        width=4.521in,
        height=3.566in,
        scale only axis,
        unbounded coords=jump,
        xmin=1,
        xmax=6.5,
        xlabel style={font=\color{white!15!black}},
        xlabel={$E_b/N_0$ [dB]},
        ymode=log,
        ymin=1e-07,
        ymax=1,
        yminorticks=true,
        ylabel style={font=\color{white!15!black}},
        ylabel={TC Rejection Probability},
        axis background/.style={fill=white},
        title style={font=\bfseries},
        xmajorgrids,
        ymajorgrids,
        yminorgrids,
        legend style={at={(0.054,0.058)}, anchor=south west, legend cell align=left, align=left, draw=white!15!black}
        ]

    \draw[black, thick] (axis cs:1,1e-5) -- (axis cs:7,1e-5);
        
        \addplot [color=blue, mark=x, mark options={solid, blue}]
          table[row sep=crcr]{%
        0	1\\
        1	0.813002649097218\\
        2	0.4384\\
        3	0.0948\\
        4	0.00661405303125915\\
        5	7.6307e-05\\
        6	3.9247e-06\\
        6.5	3.1314e-06\\
        };
        \addlegendentry{$P_{\mathrm{TCrej}}$, $N_{\mathrm{it}}=100$}

        \addplot [color=blue, dashed, mark=x, mark options={solid, blue}]
          table[row sep=crcr]{%
        0	1\\
        1	0.862001564412852\\
        2	0.545000531750865\\
        3	0.161000747643103\\
        4	0.0174009285570585\\
        5	0.0005292602525926289\\
        6	4.3646e-06\\
        6.5	7.2710E-07\\
        };
        \addlegendentry{$P_{\mathrm{TCrej}}$, $N_{\mathrm{it}}=20$}
        \end{axis}
        \end{tikzpicture}
        }
        \caption{\ac{MSA}}
    \end{subfigure}
    \hfill
    \begin{subfigure}[b]{0.48\textwidth}
        \centering
        \resizebox{\textwidth}{!}{
        \begin{tikzpicture}
        \begin{axis}[%
        width=4.521in,
        height=3.566in,
        scale only axis,
        unbounded coords=jump,
        xmin=1,
        xmax=6.5,
        xlabel style={font=\color{white!15!black}},
        xlabel={$E_b/N_0$ [dB]},
        ymode=log,
        ymin=1e-07,
        ymax=1,
        yminorticks=true,
        ylabel style={font=\color{white!15!black}},
        ylabel={TC Rejection Probability},
        axis background/.style={fill=white},
        title style={font=\bfseries},
        xmajorgrids,
        ymajorgrids,
        yminorgrids,
        legend style={at={(0.054,0.058)}, anchor=south west, legend cell align=left, align=left, draw=white!15!black}
        ]

    \draw[black, thick] (axis cs:1,1e-5) -- (axis cs:7,1e-5);
        
        \addplot [color=black, mark=asterisk, mark options={solid, black}]
          table[row sep=crcr]{%
        0	1\\
        1	0.746015038278808\\
        2	0.302024561569932\\
        3	0.0517\\
        4	0.0024433956000594\\
        5	6.3329e-05\\
        6	4.896e-05\\
        6.5	4.5968e-05\\
        };
        \addlegendentry{$P_{\mathrm{TCrej}}$, $N_{\mathrm{it}}=100$}

        \addplot [color=black, dashed, mark=asterisk, mark options={solid, black}]
          table[row sep=crcr]{%
        0	1\\
        1	0.800003921247626\\
        2	0.4000059990987\\
        3	0.0719\\
        4	0.0053\\
        5	9.2499025951289e-05\\
        6	1.598e-05\\
        6.5	1.2214e-05\\
        };
        \addlegendentry{$P_{\mathrm{TCrej}}$, $N_{\mathrm{it}}=20$}
        \end{axis}
        \end{tikzpicture}
        }
        \caption{\ac{NMSA}}
    \end{subfigure}
    \caption{TC rejection probability for the three considered decoders (randomized case)}
    \label{fig:tcrejran}
\end{figure*}

Combining these probabilities for the randomized \ac{TS} at the receiving end to determine the \ac{TC} rejection probability \eqref{eq:P_TCrej}, we obtain the results shown in Fig. \ref{fig:tcrejran}.
We observe that the curves exhibit two different behaviours: in the leftmost part, $P_{\mathrm{TCrej}}$ is mainly influenced by the CER, thus assuming the typical shape of the error probability of a coded system; in the rightmost one, we notice the presence of an error floor, due to the ``flatness'' of $P_{\mathrm{nat}}$ which, for high values of $\frac{E_b}{N_0}$, becomes the dominant term.

Let us consider an hypothetical system requirement of $P_{\mathrm{TCrej}}=10^{-5}$ around the \ac{SNR} working point, say about $6$ dB. From Fig. \ref{fig:tcrejran} we notice that, if the decoder runs at most $100$ iterations, only the \ac{MSA} (and barely the \ac{LLR-SPA}) achieves the hypothetical target performance, with relatively little margin. This is due to the fact that, in the considered $E_b/N_0$ range, $P_{\mathrm{nat}}$ is basically flat and, in particular for the \ac{LLR-SPA} and \ac{NMSA} decoders, it assumes relatively high values, with respect to $P_{\mathrm{md}}$ and \ac{CER}. Reducing the maximum number of decoding iterations improves the performance in the error floor region, at least for the case of $N = 1$ we are referring to. This happens because, by reducing $N_\mathrm{it}$, $P_\mathrm{nat}$ (leading term in \eqref{eq:P_TCrej} when $E_b/N_0$ is relatively high) gets significantly smaller. In fact, when the decoder is allowed to perform a smaller maximum number of iterations, its correction capability decreases and the noisy \ac{TS} is less likely to be corrected into a codeword. However, as we show in Section \ref{sec:tcrej}-\ref{subsec:manycods}, when $N$ increases,  $P_\mathrm{LDPC}$ might become the leading term even for relatively high values of $E_b/N_0$, thus canceling the beneficial effect of reducing the maximum number of decoding iterations.  In the next section, instead, we propose a different solution, which always improves the overall system performance.

\subsection{TC rejection probability in the de-randomized case}

In order to improve the performance discussed in Section \ref{sec:numres}-\ref{subsec:randotc}, let us now consider the randomized \ac{TS}, which (in hexadecimal) is 
\begin{equation}
\mathbf{t}'=\mathrm{AA6C\,CB0C\,C243\,AC5F\,39DC\,7AF4\,640B\,5D95},    
\end{equation}
to be appended to the data in the \ac{CLTU} encapsulation phase. This way, at the receiving end, it gets de-randomized. The component probabilities (at the receiving end) in the de-randomized case are shown in Fig.s \ref{fig:component_comp_notrand},  \ref{fig:component_comp_notrand_ms}, and  \ref{fig:component_comp_notrand_nms} for the \ac{LLR-SPA}, the \ac{MSA}, and the \ac{NMSA}, respectively. We observe that $P_{\mathrm{nat}}$ is much smaller in all these cases, with respect to the randomized case. A more thorough comparison of the $P_{\mathrm{nat}}$ behaviors is illustrated in Fig. \ref{fig:combined_pnats}.  It is remarkable that when the decoder processes the noisy de-randomized \ac{TS} instead of the randomized one, independent of the chosen algorithm, $P_\mathrm{nat}$ is always smaller than the corresponding probability in the randomized case, for all the considered values of $E_b/N_0$. It is also noticeable that, in some cases, the probability of not-acknowledged termination increases with higher values of \(E_b/N_0\). Although this may seem counterintuitive, it is important to remember that we are not transmitting a codeword. Therefore, it is possible that higher levels of noise, when added to the \ac{TS}, may on average move the received signal closer to the nearest codewords, with respect to lower levels of noise, this way increasing $P_{\mathrm{nat}}$.  Some further insights on this issue are provided in Appendix \ref{app:Hystograms}.

 \begin{figure*}
    \centering
    \begin{subfigure}[b]{0.48\textwidth}
        \centering
        \resizebox{\textwidth}{!}{\begin{tikzpicture}

\begin{axis}[%
width=4.521in,
height=3.566in,
at={(0.758in,0.481in)},
scale only axis,
unbounded coords=jump,
xmin=1,
xmax=6.5,
xlabel style={font=\color{white!15!black}},
xlabel={$E_b/N_0$ [dB]},
ymode=log,
ymin=1e-12,
ymax=1,
yminorticks=true,
ylabel style={font=\color{white!15!black}},
ylabel={Probability},
axis background/.style={fill=white},
title style={font=\bfseries},
xmajorgrids,
ymajorgrids,
yminorgrids,
legend style={at={(0.551,0.059)}, anchor=south west, legend cell align=left, align=left, draw=white!15!black}
]
\addplot [color=red, mark=square, mark options={solid, red}]
  table[row sep=crcr]{%
0   0.000381413109508899\\
1	1.05063337376698e-05\\
2	8.84987106930524e-08\\
3	1.75540471047952e-10\\
4	5.93969318174459e-14\\
5	1.33226762955019e-15\\
6	1.63757896132211e-15\\
7	1.04083408558608e-15\\
};
\addlegendentry{$P_\mathrm{md}$}

\addplot [color=blue, mark=x, mark options={solid, blue}]
  table[row sep=crcr]{%
1 7.576E-01\\
2 3.155E-01\\
3 5.173E-02\\
4 2.203E-03\\
5 2.358E-05\\
6 1.34E-07\\
6.5 7.69E-09\\
};
\addlegendentry{CER, $N_{\mathrm{it}}=100$}

\addplot [color=blue, dashed, mark=x, mark options={solid, blue}]
  table[row sep=crcr]{%
1 7.75E-01\\
2 3.268E-01\\
3 6.849E-02\\
4 5.986E-03\\
5 1.773E-04\\
6 1.89E-06\\
6.5 1.246E-07\\
7 8.336E-09\\
};
\addlegendentry{CER, $N_{\mathrm{it}}=20$}

\addplot [color=black, mark=o, mark options={solid, black}]
  table[row sep=crcr]{%
0	2.00000466667756e-06\\
1	1.098E-05\\ 
2	2.813E-05\\ 
3	4.026E-05\\ 
4	3.396E-05\\ 
5	1.711e-05\\
6	6.198e-06\\
7	1.457E-06 \\
};
\addlegendentry{$P_\mathrm{nat}$, $N_{\mathrm{it}}=100$}

\addplot [color=black, dashed, mark=o, mark options={solid, black}]
  table[row sep=crcr]{%
1	5.073E-06\\  
2	1.642E-05 \\ 
3	2.02E-05\\ 
4	1.167E-05 \\ 
5 3.902E-06\\ 
6 1.08E-06 \\ 
6.5 4.534E-07\\ 
7 2.342E-07\\
};
\addlegendentry{$P_\mathrm{nat}$, $N_{\mathrm{it}}=20$}

\end{axis}

\begin{axis}[%
width=5.833in,
height=4.375in,
at={(0in,0in)},
scale only axis,
xmin=0,
xmax=1,
ymin=0,
ymax=1,
axis line style={draw=none},
ticks=none,
axis x line*=bottom,
axis y line*=left
]
\end{axis}
\end{tikzpicture}%
}
        \caption{\ac{LLR-SPA}}
        \label{fig:component_comp_notrand}
    \end{subfigure}
    \hfill
    \begin{subfigure}[b]{0.48\textwidth}
        \centering
        \resizebox{\textwidth}{!}{\begin{tikzpicture}

\begin{axis}[%
width=4.521in,
height=3.566in,
at={(0.758in,0.481in)},
scale only axis,
unbounded coords=jump,
xmin=1,
xmax=6.5,
xlabel style={font=\color{white!15!black}},
xlabel={$E_b/N_0$ [dB]},
ymode=log,
ymin=1e-12,
ymax=1,
yminorticks=true,
ylabel style={font=\color{white!15!black}},
ylabel={Probability},
axis background/.style={fill=white},
title style={font=\bfseries},
xmajorgrids,
ymajorgrids,
yminorgrids,
legend style={at={(0.551,0.059)}, anchor=south west, legend cell align=left, align=left, draw=white!15!black}
]
\addplot [color=red, mark=square, mark options={solid, red}]
  table[row sep=crcr]{%
0   0.000381413109508899\\
1	1.05063337376698e-05\\
2	8.84987106930524e-08\\
3	1.75540471047952e-10\\
4	5.93969318174459e-14\\
5	1.33226762955019e-15\\
6	1.63757896132211e-15\\
7	1.04083408558608e-15\\
};
\addlegendentry{$P_{\mathrm{md}}$}

\addplot [color=blue, mark=x, mark options={solid, blue}]
  table[row sep=crcr]{%
1 8.13E-01\\
2 4.384E-01\\
3 9.477E-02\\
4 6.633E-03\\
5 7.216E-05\\
6 1.89E-07\\
6.5 8.22E-09  \\
};
\addlegendentry{CER, , $N_{\mathrm{it}}=100$}

\addplot [color=blue, dashed, mark=x, mark options={solid, blue}]
  table[row sep=crcr]{%
1 8.62E-01 \\
2 5.45E-01\\
3 1.61E-01\\
4 1.739E-02\\
5 5.284E-04 \\
6 3.7E-06\\
6.5 2.18E-07\\
7 1.052E-08\\
};
\addlegendentry{CER, $N_{\mathrm{it}}=20$}

\addplot [color=black, mark=o, mark options={solid, black}]
  table[row sep=crcr]{%
0	4.33335355564993e-06\\
1	2.746e-06\\
2	2.474e-06\\
3	2.128e-06\\
4	1.58e-06\\
5	1.05333444444481e-06\\
6	7.00000100000100e-07\\
7	3.33333444444481e-07\\
};
\addlegendentry{$P_{\mathrm{nat}}$, $N_{\mathrm{it}}=100$}

\addplot [color=black, dashed, mark=o, mark options={solid, black}]
  table[row sep=crcr]{%
1 7.025E-07\\
2 5.99E-07 \\ 
3 4.932E-07\\ 
4 3.4E-07\\ 
5 2.097e-07\\
6 1.043e-07\\
6.5 6.055e-08\\
7 4.12E-08\\ 
};
\addlegendentry{$P_{\mathrm{nat}}$, $N_{\mathrm{it}}=20$}

\end{axis}

\begin{axis}[%
width=5.833in,
height=4.375in,
at={(0in,0in)},
scale only axis,
xmin=0,
xmax=1,
ymin=0,
ymax=1,
axis line style={draw=none},
ticks=none,
axis x line*=bottom,
axis y line*=left
]
\end{axis}
\end{tikzpicture}%
}
        \caption{\ac{MSA}}
        \label{fig:component_comp_notrand_ms}
    \end{subfigure}
    
    \begin{subfigure}[b]{0.48\textwidth}
        \centering
        \resizebox{\textwidth}{!}{\begin{tikzpicture}

\begin{axis}[%
width=4.521in,
height=3.566in,
at={(0.758in,0.481in)},
scale only axis,
unbounded coords=jump,
xmin=1,
xmax=6.5,
xlabel style={font=\color{white!15!black}},
xlabel={$E_b/N_0$ [dB]},
ymode=log,
ymin=1e-12,
ymax=1,
yminorticks=true,
ylabel style={font=\color{white!15!black}},
ylabel={Probability},
axis background/.style={fill=white},
title style={font=\bfseries},
xmajorgrids,
ymajorgrids,
yminorgrids,
legend style={at={(0.551,0.059)}, anchor=south west, legend cell align=left, align=left, draw=white!15!black}
]
\addplot [color=red, mark=square, mark options={solid, red}]
  table[row sep=crcr]{%
0   0.000381413109508899\\
1	1.05063337376698e-05\\
2	8.84987106930524e-08\\
3	1.75540471047952e-10\\
4	5.93969318174459e-14\\
5	1.33226762955019e-15\\
6	1.63757896132211e-15\\
7	1.04083408558608e-15\\
};
\addlegendentry{$P_{\mathrm{md}}$}

\addplot [color=blue, mark=x, mark options={solid, blue}]
  table[row sep=crcr]{%
1 7.46E-01 \\
2 3.02E-01\\
3 5.17E-02\\
4 2.4E-03\\
5 1.6E-05 \\
6 5.99E-08\\
6.5 2.973E-09\\
};
\addlegendentry{CER, $N_{\mathrm{it}}=100$}

\addplot [color=blue, dashed,  mark=x, mark options={solid, blue}]
  table[row sep=crcr]{%
1 8E-01 \\
2 4E-01\\
3 7.184E-02\\
4 5.33E-03\\
5 8.05E-05 \\
6 3.8E-07\\
6.5 2.436E-08\\
};
\addlegendentry{CER, $N_{\mathrm{it}}=20$}

\addplot [color=black, mark=o, mark options={solid, black}]
  table[row sep=crcr]{%
1 3.086E-05\\	
2 3.014E-05\\ 
3 2.669E-05\\	
4	2.645E-05 \\ 
5	2.469E-05\\ 
6	1.664E-05\\ 
7	1.019E-05\\ 
};
\addlegendentry{$\mathrm{P_{nat}}$, $N_{\mathrm{it}}=100$}

\addplot [color=black, mark=o, dashed, mark options={solid, black}]
  table[row sep=crcr]{%
1	7.843E-06 \\ 
2	7.483E-06  \\ 
3	6.286E-06   \\ 
4	5.292E-06\\ 
5	3.93E-06\\ 
6	2.72E-06 \\ 
7	1.4E-06\\ 
8 6.894E-07\\
};
\addlegendentry{$\mathrm{P_{nat}}$, $N_{\mathrm{it}}=20$}

\end{axis}

\begin{axis}[%
width=5.833in,
height=4.375in,
at={(0in,0in)},
scale only axis,
xmin=0,
xmax=1,
ymin=0,
ymax=1,
axis line style={draw=none},
ticks=none,
axis x line*=bottom,
axis y line*=left
]
\end{axis}
\end{tikzpicture}%
}
        \caption{\ac{NMSA}}
        \label{fig:component_comp_notrand_nms}
    \end{subfigure}
    
    \caption{Comparison of \ac{TC} rejection probability's leading components for different algorithms using the proposed new solution (de-randomized case)}
    \label{fig:combined_notrand}
\end{figure*}
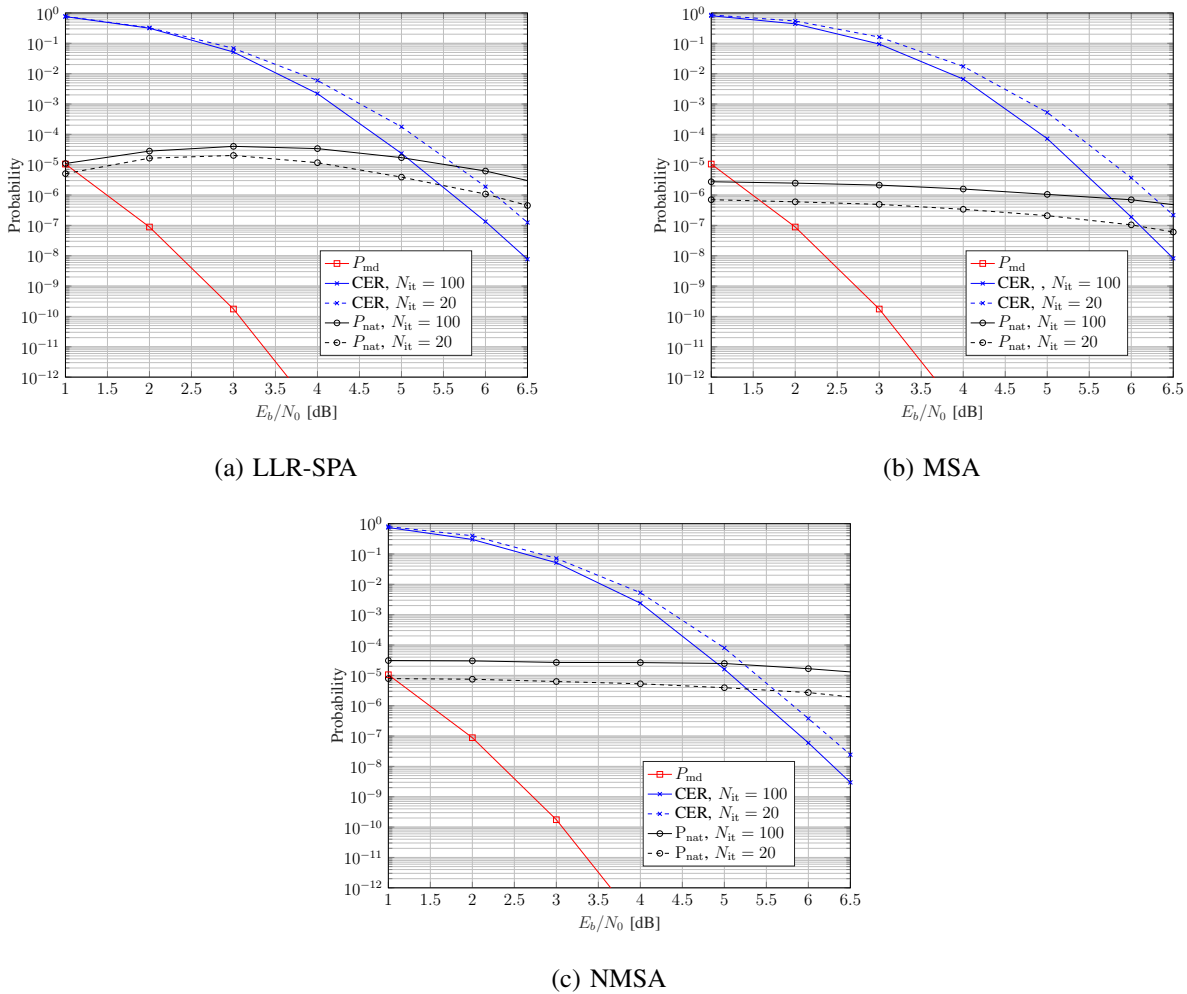
   
\begin{figure*}
    \centering
    \begin{subfigure}[b]{0.48\textwidth}
        \centering
        \resizebox{\textwidth}{!}{\begin{tikzpicture}
\begin{axis}[%
width=4.521in,
height=3.566in,
at={(0.758in,0.481in)},
scale only axis,
unbounded coords=jump,
xmin=1,
xmax=9,
xlabel style={font=\color{white!15!black}},
xlabel={$E_b/N_0$ [dB]},
ymode=log,
ymin=5e-8,
ymax=1e-04,
yminorticks=true,
ylabel style={font=\color{white!15!black}},
ylabel={Probability of Not-Acknowledged Termination},
axis background/.style={fill=white},
xmajorgrids,
ymajorgrids,
yminorgrids,
legend style={at={(0,0)}, anchor=south west, legend cell align=left, align=left, draw=white!15!black}
]

\addplot [color=orange, mark=o, mark options={solid, orange}]
  table[row sep=crcr]{%
0	2.00000466667756e-06\\ 
1	1.563E-05\\ 
2	3.902E-05\\ 
3	5.458e-05\\
4	4.383e-05\\
5	3.010e-05\\
6	1.628e-05\\
7	6.90e-06\\ 
8   2.711e-06\\
};
\addlegendentry{Rand., $N_{\mathrm{it}}=100$}

\addplot [color=blue, mark=o, mark options={solid, blue}]
  table[row sep=crcr]{%
1	7.91E-06\\ 
2	2.168E-05\\ 
3	3.175E-05\\ 
4	1.591E-05\\ 
5	8.39E-06 \\ 
6 3.713E-06 \\ 
7 1.306E-06\\ 
};
\addlegendentry{Rand., $N_{\mathrm{it}}=20$}

\addplot [color=orange, mark=x, mark options={solid, orange}]
  table[row sep=crcr]{%
0	2.00000466667756e-06\\
1	1.098E-05\\ 
2	2.813E-05\\ 
3	4.026E-05\\ 
4	3.396E-05\\ 
5	1.711e-05\\
6	6.198e-06\\
7	1.457E-06 \\
};
\addlegendentry{De-Rand., $N_{\mathrm{it}}=100$}

\addplot [color=blue, mark=x, mark options={solid, blue}]
  table[row sep=crcr]{%
1	5.073E-06\\  
2	1.642E-05 \\ 
3	2.02E-05\\ 
4	1.167E-05 \\ 
5 3.902E-06\\ 
6 1.08E-06 \\ 
6.5 4.534E-07\\ 
};
\addlegendentry{De-Rand., $N_{\mathrm{it}}=20$}

\end{axis}
\end{tikzpicture}}
        \caption{\ac{LLR-SPA}}
        \label{fig:nat_notrand}
    \end{subfigure}
    \hfill
    \begin{subfigure}[b]{0.48\textwidth}
        \centering
        \resizebox{\textwidth}{!}{\begin{tikzpicture}
\begin{axis}[%
width=4.521in,
height=3.566in,
at={(0.758in,0.481in)},
scale only axis,
unbounded coords=jump,
xmin=1,
xmax=9,
xlabel style={font=\color{white!15!black}},
xlabel={$E_b/N_0$ [dB]},
ymode=log,
ymin=5e-8,
ymax=1e-04,
yminorticks=true,
ylabel style={font=\color{white!15!black}},
ylabel={Probability of Not-Acknowledged Termination},
axis background/.style={fill=white},
xmajorgrids,
ymajorgrids,
yminorgrids,
legend style={at={(0.6,0.75)}, anchor=south west, legend cell align=left, align=left, draw=white!15!black}
]

\addplot [color=red, mark=o, mark options={solid, red}]
  table[row sep=crcr]{%
0	5.33336355572682e-06\\ 
1	3.65617077534951e-06\\
2	3.367E-06\\ 
3	3.55065357057806e-06\\
4	4.07575724105145e-06\\
5	4.147e-06\\
6	3.73572760598474e-06\\
7	2.51071649119939e-06\\
8 1.529E-06\\
};
\addlegendentry{Rand., $N_{\mathrm{it}}=100$}

\addplot [color=cyan, mark=o, mark options={solid, cyan}]
  table[row sep=crcr]{%
1	9.143E-07 \\ 
2 9.38E-07\\ 
3 9.242E-07\\
4 1.003E-06\\ 
5 8.639E-07 \\ 
6 6.646E-07 \\ 
7 3.536E-07\\ 
};
\addlegendentry{Rand., $N_{\mathrm{it}}=20$}

\addplot [color=red, mark=x, mark options={solid, red}]
  table[row sep=crcr]{%
0	4.33335355564993e-06\\
1	2.746e-06\\ 
2	2.474e-06\\
3	2.128E-06\\
4	1.58e-06\\
5	1.05333444444481e-06\\
6	7.00000100000100e-07\\
7	3.33333444444481e-07\\
};
\addlegendentry{De-Rand., $N_{\mathrm{it}}=100$}

\addplot [color=cyan,  mark=x, mark options={solid, cyan}]
  table[row sep=crcr]{%
1 7.025E-07\\
2 5.99E-07 \\ 
3 4.932E-07\\ 
4 3.4E-07\\ 
5 2.097e-07\\
6 1.043e-07\\
6.5 6.055E-08\\
};
\addlegendentry{De-Rand., $N_{\mathrm{it}}=20$}

\end{axis}
\end{tikzpicture}}
        \caption{\ac{MSA}}
        \label{fig:nat_notrand_ms}
    \end{subfigure}
    
    \begin{subfigure}[b]{0.48\textwidth}
        \centering
        \resizebox{\textwidth}{!}{\begin{tikzpicture}
\begin{axis}[%
width=4.521in,
height=3.566in,
at={(0.758in,0.481in)},
scale only axis,
unbounded coords=jump,
xmin=1,
xmax=9,
xlabel style={font=\color{white!15!black}},
xlabel={$E_b/N_0$ [dB]},
ymode=log,
ymin=5e-8,
ymax=1e-04,
yminorticks=true,
ylabel style={font=\color{white!15!black}},
ylabel={Probability of Not-Acknowledged Termination},
axis background/.style={fill=white},
xmajorgrids,
ymajorgrids,
yminorgrids,
legend style={at={(0,0)}, anchor=south west, legend cell align=left, align=left, draw=white!15!black}
]

\addplot [color=black, mark=o, mark options={solid, black}]
  table[row sep=crcr]{%
1	3.81E-05 \\ 
2	3.43E-05 \\ 
3	3.847E-05 \\
4	4.24E-05\\ 
5	4.733E-05\\ 
6	4.89E-05\\ 
7	4.303E-05\\ 
8 2.685E-05\\ 
9 1.317E-05\\ 
};
\addlegendentry{Rand., $N_{\mathrm{it}}=100$}

\addplot [color=purple, mark=o, mark options={solid, purple}]
  table[row sep=crcr]{%
1	9.758E-06\\ 
2	9.728E-06\\ 
3	1.027E-05\\ 
4	1.056E-05\\ 
5	1.2E-05\\ 
6	1.183E-05\\ 
7	7.733E-06\\ 
8 4.750E-06\\ 
9  2.085E-06\\ 
};
\addlegendentry{Rand., $N_{\mathrm{it}}=20$}

\addplot [color=black, mark=x, mark options={solid, black}]
  table[row sep=crcr]{%
1 3.086E-05\\	
2 3.014E-05\\ 
3 2.669E-05\\	
4	2.645E-05 \\ 
5	2.469E-05\\ 
6	1.664E-05\\ 
7	1.019E-05\\ 
8 6.588E-06\\
9 2.641E-06\\
};
\addlegendentry{De-Rand., $N_{\mathrm{it}}=100$}

\addplot [color=purple, mark=x, mark options={solid, purple}]
  table[row sep=crcr]{%
1	7.843E-06 \\ 
2	7.483E-06  \\ 
3	6.286E-06   \\ 
4	5.292E-06\\ 
5	3.93E-06\\ 
6	2.72E-06 \\ 
7	1.4E-06\\ 
8 6.894E-07\\
};
\addlegendentry{De-Rand., $N_{\mathrm{it}}=20$}

\end{axis}
\end{tikzpicture}}
        \caption{\ac{NMSA}}
        \label{fig:nat_notrand_nms}
    \end{subfigure}
    
    \caption{Not-acknowledged rejection probabilities for different decoding algorithms, in the randomized case (standard solution) and in the de-randomized case (proposed solution)}
    \label{fig:combined_pnats}
\end{figure*}
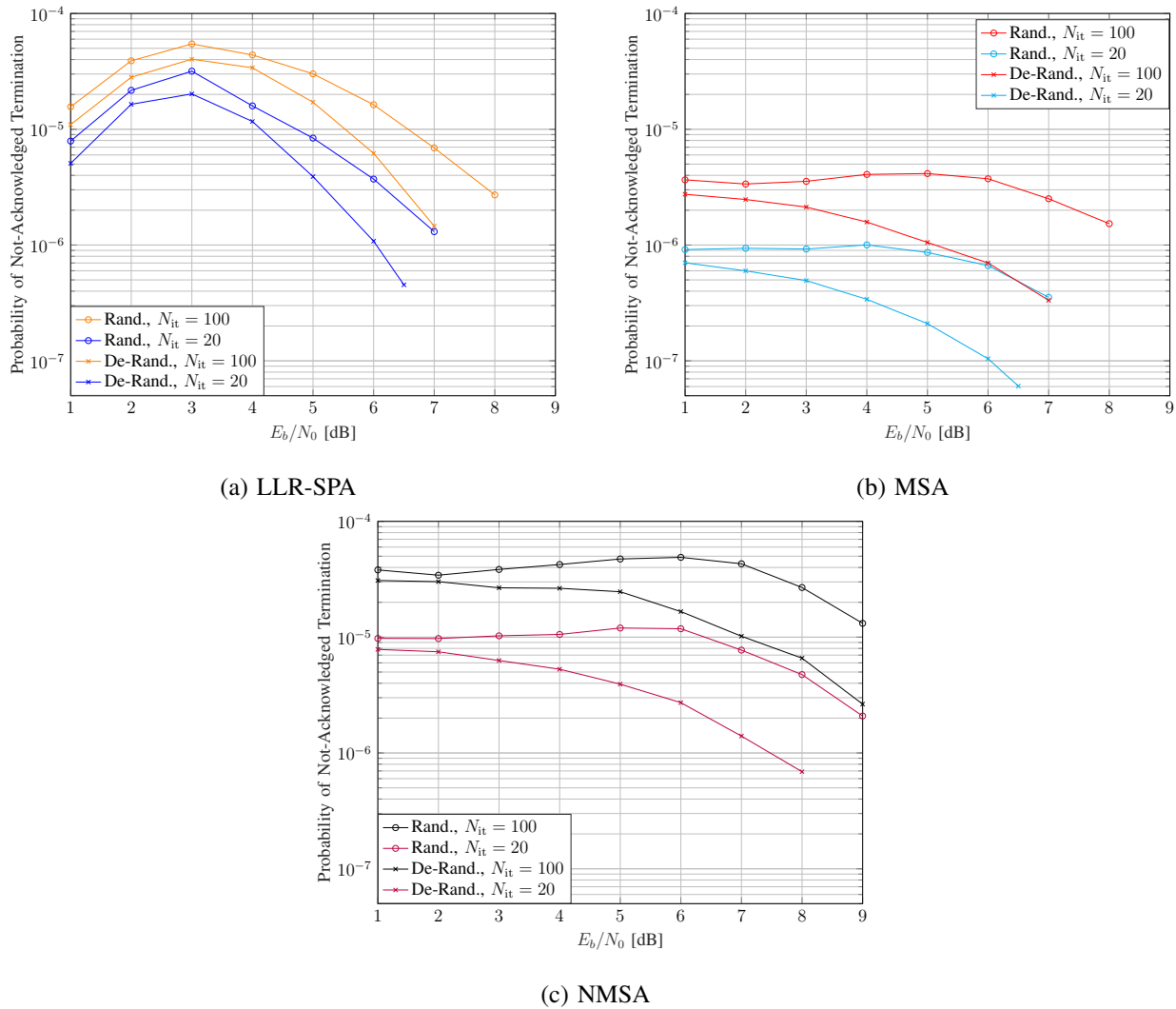

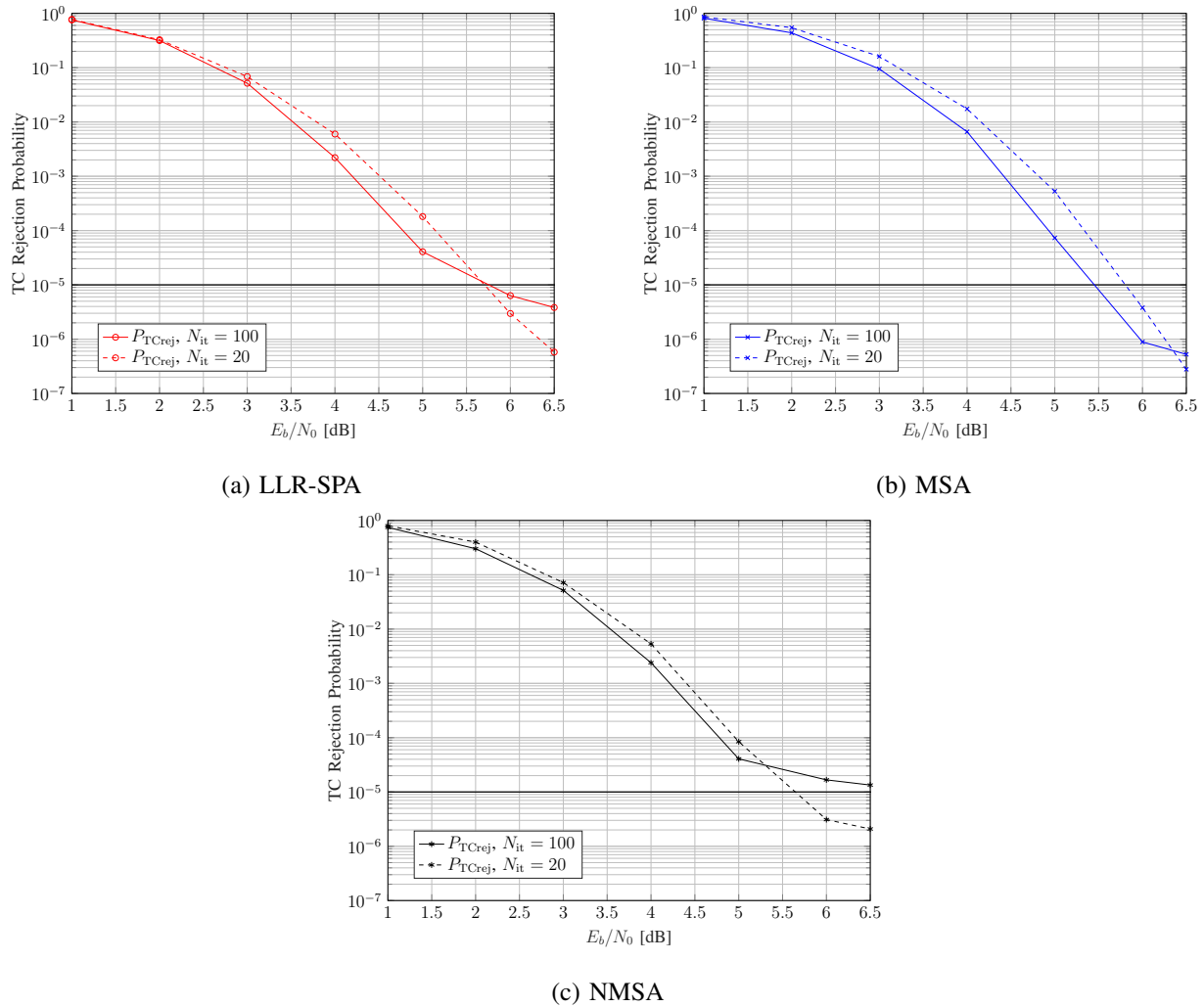
\begin{figure*}
    \centering
    \begin{subfigure}[b]{0.48\textwidth}
        \centering
        \resizebox{\textwidth}{!}{
        \begin{tikzpicture}
        \begin{axis}[%
        width=4.521in,
        height=3.566in,
        scale only axis,
        unbounded coords=jump,
        xmin=1,
        xmax=6.5,
        xlabel style={font=\color{white!15!black}},
        xlabel={$E_b/N_0$ [dB]},
        ymode=log,
        ymin=1e-07,
        ymax=1,
        yminorticks=true,
        ylabel style={font=\color{white!15!black}},
        ylabel={TC Rejection Probability},
        axis background/.style={fill=white},
        title style={font=\bfseries},
        xmajorgrids,
        ymajorgrids,
        yminorgrids,
        legend style={at={(0.054,0.058)}, anchor=south west, legend cell align=left, align=left, draw=white!15!black}
        ]

    \draw[black, thick] (axis cs:1,1e-5) -- (axis cs:7,1e-5);
        
        \addplot [color=red, mark=o, mark options={solid, red}]
          table[row sep=crcr]{%
        0	1\\
        1	0.758\\
        2	0.3155\\ 
        3	0.0517\\ 
        4	0.0022\\ 
        5	4.07E-05\\ 
        6	6.3320E-06\\ 
        6.5 3.8352E-06\\
        };
        \addlegendentry{$P_{\mathrm{TCrej}}$, $N_{\mathrm{it}}=100$}

        \addplot [color=red, dashed, mark=o, mark options={solid, red}]
          table[row sep=crcr]{%
        0	1\\
        1	0.775003500163153\\
        2	0.327010894858674\\
        3	0.0685192822135125\\
        4	0.00600093411005918\\
        5	0.0001812009113231289\\
        6	2.97e-06\\
        6.5 5.78E-07\\
        };
        \addlegendentry{$P_{\mathrm{TCrej}}$, $N_{\mathrm{it}}=20$}
        \end{axis}
        \end{tikzpicture}
        }
        \caption{\ac{LLR-SPA}}
    \end{subfigure}
    \hfill
    \begin{subfigure}[b]{0.48\textwidth}
        \centering
        \resizebox{\textwidth}{!}{
        \begin{tikzpicture}
        \begin{axis}[%
        width=4.521in,
        height=3.566in,
        scale only axis,
        unbounded coords=jump,
        xmin=1,
        xmax=6.5,
        xlabel style={font=\color{white!15!black}},
        xlabel={$E_b/N_0$ [dB]},
        ymode=log,
        ymin=1e-07,
        ymax=1,
        yminorticks=true,
        ylabel style={font=\color{white!15!black}},
        ylabel={TC Rejection Probability},
        axis background/.style={fill=white},
        title style={font=\bfseries},
        xmajorgrids,
        ymajorgrids,
        yminorgrids,
        legend style={at={(0.054,0.058)}, anchor=south west, legend cell align=left, align=left, draw=white!15!black}
        ]

    \draw[black, thick] (axis cs:1,1e-5) -- (axis cs:7,1e-5);
        
        \addplot [color=blue, mark=x, mark options={solid, blue}]
          table[row sep=crcr]{%
        0	1\\
        1	0.813002456489242\\
        2	0.438400\\
        3	0.0948\\
        4	0.0066\\
        5	7.3213e-05\\
        6	8.89e-07\\
        6.5 5.2472E-07\\
        };
        \addlegendentry{$P_{\mathrm{TCrej}}$, $N_{\mathrm{it}}=100$}

        \addplot [color=blue, dashed, mark=x, mark options={solid, blue}]
          table[row sep=crcr]{%
        0	1\\
        1	0.862001542609082\\
        2	0.5450\\
        3	0.161000446801103\\
        4	0.0174003350666585\\
        5	0.00052861\\
        6	3.8043e-06\\
        6.5 2.7855E-07\\
        };
        \addlegendentry{$P_{\mathrm{TCrej}}$, $N_{\mathrm{it}}=20$}
        \end{axis}
        \end{tikzpicture}
        }
        \caption{\ac{MSA}}
    \end{subfigure}
    \hfill
    \begin{subfigure}[b]{0.48\textwidth}
        \centering
        \resizebox{\textwidth}{!}{
        \begin{tikzpicture}
        \begin{axis}[%
        width=4.521in,
        height=3.566in,
        scale only axis,
        unbounded coords=jump,
        xmin=1,
        xmax=6.5,
        xlabel style={font=\color{white!15!black}},
        xlabel={$E_b/N_0$ [dB]},
        ymode=log,
        ymin=1e-07,
        ymax=1,
        yminorticks=true,
        ylabel style={font=\color{white!15!black}},
        ylabel={TC Rejection Probability},
        axis background/.style={fill=white},
        title style={font=\bfseries},
        xmajorgrids,
        ymajorgrids,
        yminorgrids,
        legend style={at={(0.054,0.058)}, anchor=south west, legend cell align=left, align=left, draw=white!15!black}
        ]

    \draw[black, thick] (axis cs:1,1e-5) -- (axis cs:7,1e-5);
        
        \addplot [color=black, mark=asterisk, mark options={solid, black}]
          table[row sep=crcr]{%
        0	1\\
        1	0.746011177519371\\
        2	0.302020722570272\\
        3	0.0517\\
        4	0.0024\\
        5	4.069e-05\\
        6	1.67e-05\\
        6.5 1.3418E-05\\
        };
        \addlegendentry{$P_{\mathrm{TCrej}}$, $N_{\mathrm{it}}=100$}

        \addplot [color=black, dashed, mark=asterisk, mark options={solid, black}]
          table[row sep=crcr]{%
        0	1\\
        1	0.800003557251450\\
        2	0.400004913098796\\
        3	0.0718061912569355\\
        4	0.00533642556825922\\
        5	8.443e-05\\
        6	3.1e-06\\
        6.5 2.0844E-06\\
        };
        \addlegendentry{$P_{\mathrm{TCrej}}$, $N_{\mathrm{it}}=20$}
        \end{axis}
        \end{tikzpicture}
        }
        \caption{\ac{NMSA}}
    \end{subfigure}
    \caption{TC rejection probability using different decoders for the proposed solution (de-randomized case)}
    \label{fig:tcrej_nr}
\end{figure*}

The overall \ac{TC} rejection probability is shown in Fig. \ref{fig:tcrej_nr}. If we compare Fig.s \ref{fig:tcrejran} and \ref{fig:tcrej_nr} (with the help of the horizontal line representing the hypothetical system requirement), we observe that the performance in the de-randomized case is always better than that in the randomized (and thus standard) one, for all the considered decoders and maximum number of decoding iterations. Moreover, this happens not only around the considered working point, but for all values of $E_b/N_0$. This is an obvious consequence of the results in Fig. \ref{fig:combined_pnats}, since the probability of missed detection of the start sequence and $P_{\mathrm{LDPC}}$ do not change, if the \acp{TS} changes. We remark that the considered target performance is reached in all cases around the working point, except for the \ac{NMSA}-based decoder running $100$ iterations.

It is also remarkable that, in both randomized and de-randomized cases, the performance shows an error floor, especially when the maximum number of decoding iterations is set to $100$, and when either the \ac{LLR-SPA} or the \ac{NMSA} are employed. As apparent from  Fig.s \ref{fig:combined} and \ref{fig:combined_notrand}, where we show the single leading components of $P_{\mathrm{TCrej}}$, the error floor is always due to the not-acknowledged termination probability $P_{\mathrm{nat}}$ since, for relatively large values of $E_b/N_0$, the values of $P_{\mathrm{md}}$ and \ac{CER} decrease quite rapidly with $E_b/N_0$, and thus have a negligible impact in \eqref{eq:P_TCrej}.

\subsection{Decoder convergence analysis on the noisy TS}\label{subsec:decodingsuccesses}

Since the results presented in the previous subsection are worse for the randomized (standard) case than for the de-randomized one, let us analyze the rate at which the considered decoders converge to a codeword when fed with the noisy \ac{TS}, running $100$ decoding iterations.
The latter choice is motivated by the fact that, since the probability of not-acknowledged termination is larger than for $20$ decoding iterations, we expect to get a larger decoder convergence rate. 
For such a purpose, we count the number of times the decoder converges to a generic codeword, starting from the noisy \ac{TS}, by considering $3\,000\,000$ decoding instances\footnote{In this case, we run a fixed number of transmissions, rather than transmitting until encountering a fixed number of decoding successes (as done, for example, in Section \ref{sec:numres}-\ref{subsec:randotc}). As shown in Appendix \ref{app:Hystograms}, this allows a fair comparison of the decoding success rate for different values of $E_b/N_0$.}, which ensures a good statistical confidence. The codewords the decoder converges to are univocally identified by a codeword index in the Fig.s, ranging between $1$ and $1384$. The simulated values of $E_b/N_0$ range from $0$ to $7$ dB, with a step of $1$ dB, and the shown numbers collect the decoding successes for the whole range of $E_b/N_0$ we have considered. This choice follows from the fact that we are interested in studying the anomalous behaviors independent of the value of the signal-to-noise ratio. On the other hand, in Appendix \ref{app:Hystograms} we will separate the events and we will show the number of decoding successes, having \ac{TS} in input, as a function of the signal-to-noise ratio. 

\begin{figure}
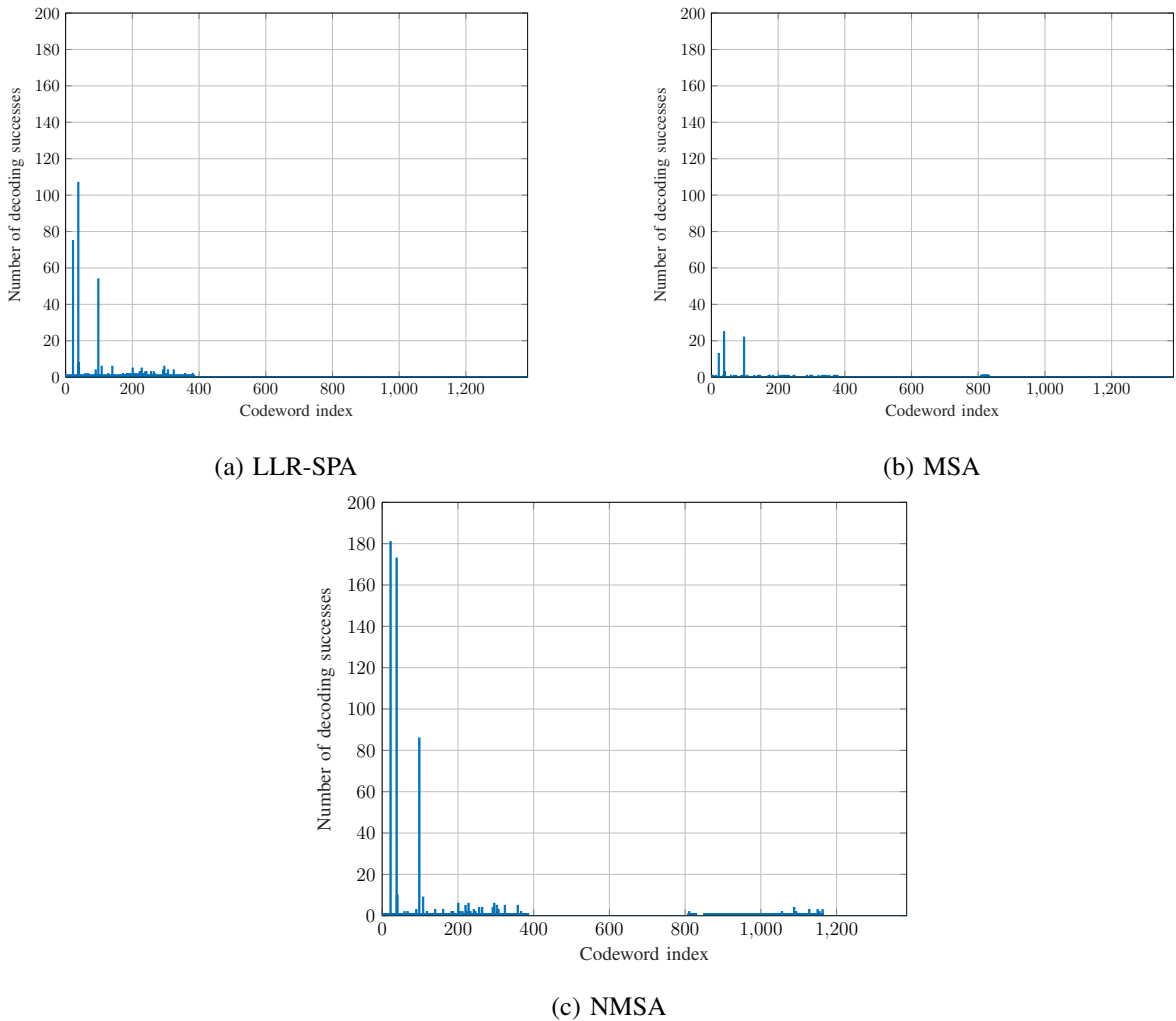

    \centering
    \begin{subfigure}[b]{0.48\textwidth}
        \centering
        \resizebox{\textwidth}{!}{\input{Images/SPALLR_RandOccur}}
        \caption{\ac{LLR-SPA}}
        \label{fig:spallr_rand}
    \end{subfigure}
    \hfill
    \begin{subfigure}[b]{0.48\textwidth}
        \centering
        \resizebox{\textwidth}{!}{\input{Images/MinSum_rand_Occur}}
        \caption{\ac{MSA}}
        \label{fig:ms_rand}
    \end{subfigure}
    
    \begin{subfigure}[b]{0.48\textwidth}
        \centering
        \resizebox{\textwidth}{!}{\input{Images/NMS_rand_Occ}}
        \caption{\ac{NMSA}}
        \label{fig:nms_rand}
    \end{subfigure}
    
    \caption{Analysis of decoding successes of noisy \ac{TS} for different decoding algorithms in the randomized case}
    \label{fig:combined_occ_rand}
\end{figure}

Observing Fig.s \ref{fig:spallr_rand}, \ref{fig:ms_rand}, and \ref{fig:nms_rand}, obtained with the three different decoders, it is evident that each decoder very frequently converged from the noisy \ac{TS} to one of three codewords, identified by the codeword indexes $22$, $38$, and $98$. 
We have verified that the Hamming distance between the \ac{TS} and these codewords is $15$. Explicitly, these codewords are (in hexadecimal):
\begin{itemize}
    \item AE6C EF4C C057 BC7F 1DDC FBF4 641B 5D85
    \item AAEC 8F0C CA43 2C5F 3F58 78F4 048B 1DB5
    \item 0A4C 8B0C C34B ACDD 29DD FEF4 250B 5D97
\end{itemize}
Then, we can say there is a ``polarization'' in the \ac{LLR-SPA}, \ac{MSA}, and \ac{NMSA} decoders towards these three codewords which, more than others, compromise the system's performance. The specific number of decoding successes justify the fact that the \ac{MSA}-based decoder has the best performance in terms of probability of not-acknowledged termination, followed by the \ac{LLR-SPA}-based decoder, and then by the \ac{NMSA}-based decoder (see Fig. \ref{fig:combined_pnats}). 

\begin{figure}
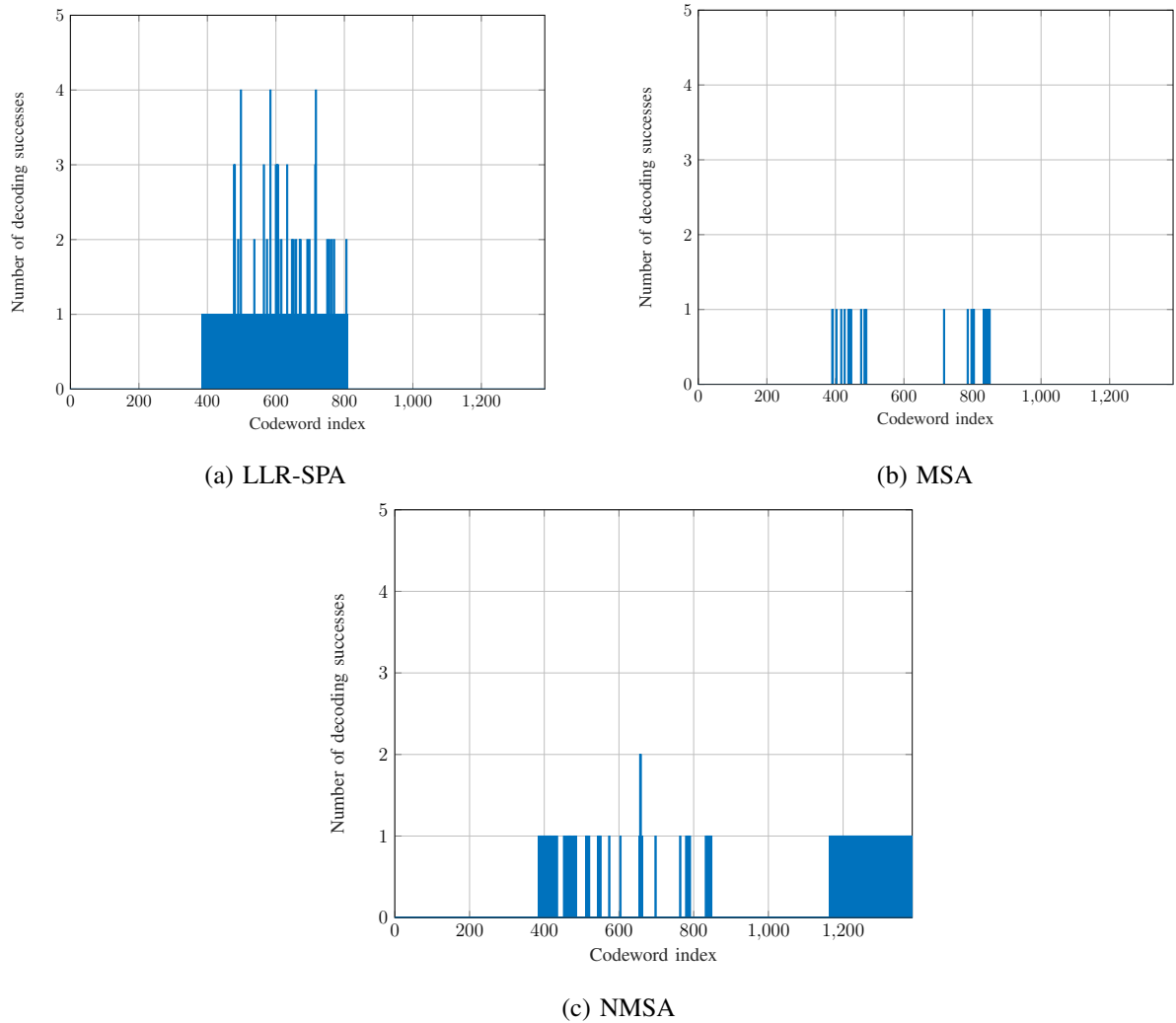

    \centering
    \begin{subfigure}[b]{0.44\textwidth}
        \centering
        \resizebox{\textwidth}{!}{\input{Images/SPALLRoccnotrand}}
        \caption{\ac{LLR-SPA}}
        \label{fig:spallr_notrand}
    \end{subfigure}
    \hfill
    \begin{subfigure}[b]{0.50\textwidth}
        \centering
        \resizebox{\textwidth}{!}{\input{Images/MSnotrandOcc}}
        \caption{\ac{MSA}}
        \label{fig:ms_notrand}
    \end{subfigure}
    
    \begin{subfigure}[b]{0.48\textwidth}
        \centering
        \resizebox{\textwidth}{!}{\input{Images/NMS_notrand_occ}}
        \caption{\ac{NMSA}}
        \label{fig:nms_notrand}
    \end{subfigure}
    
    \caption{Analysis of decoding successes of noisy \ac{TS} for different algorithms when applying the proposed solution in the de-randomized case} 
    \label{fig:combined_occ_notrand}
\end{figure}

Simulation has been repeated for the de-randomized case and results are reported in Fig. \ref{fig:combined_occ_notrand}. We observe that the ``polarization'' effect practically disappears when the solution we propose is adopted. Most importantly, the three histograms for the de-randomized case clearly show that the codewords are mistaken for the \ac{TS} significantly fewer times, indicating that they are sufficiently and almost evenly distant from the de-randomized \ac{TS}. In particular, the Hamming distance of the \ac{TS} to the closest misunderstood codewords  is $18$, rather than $15$ of the randomized case. 

To ensure that no other codewords exhibit a smaller Hamming distance than $18$ from the \ac{TS}, we employed Stern's information set decoding method \cite{stern1989method} and the technique outlined in \cite{Hu2004a} to find numerous low-weight codewords of the $(128,64)$ code. Our analysis reveals that none of these codewords has a Hamming distance smaller than $18$ from the \ac{TS}, $73$ codewords indeed differ from the \ac{TS} in  $18$ coordinates, while $3967$ codewords have a Hamming distance of $20$ from the \ac{TS}.

\subsection{Analysis for a larger number of codewords in the CLTU}
\label{subsec:manycods}

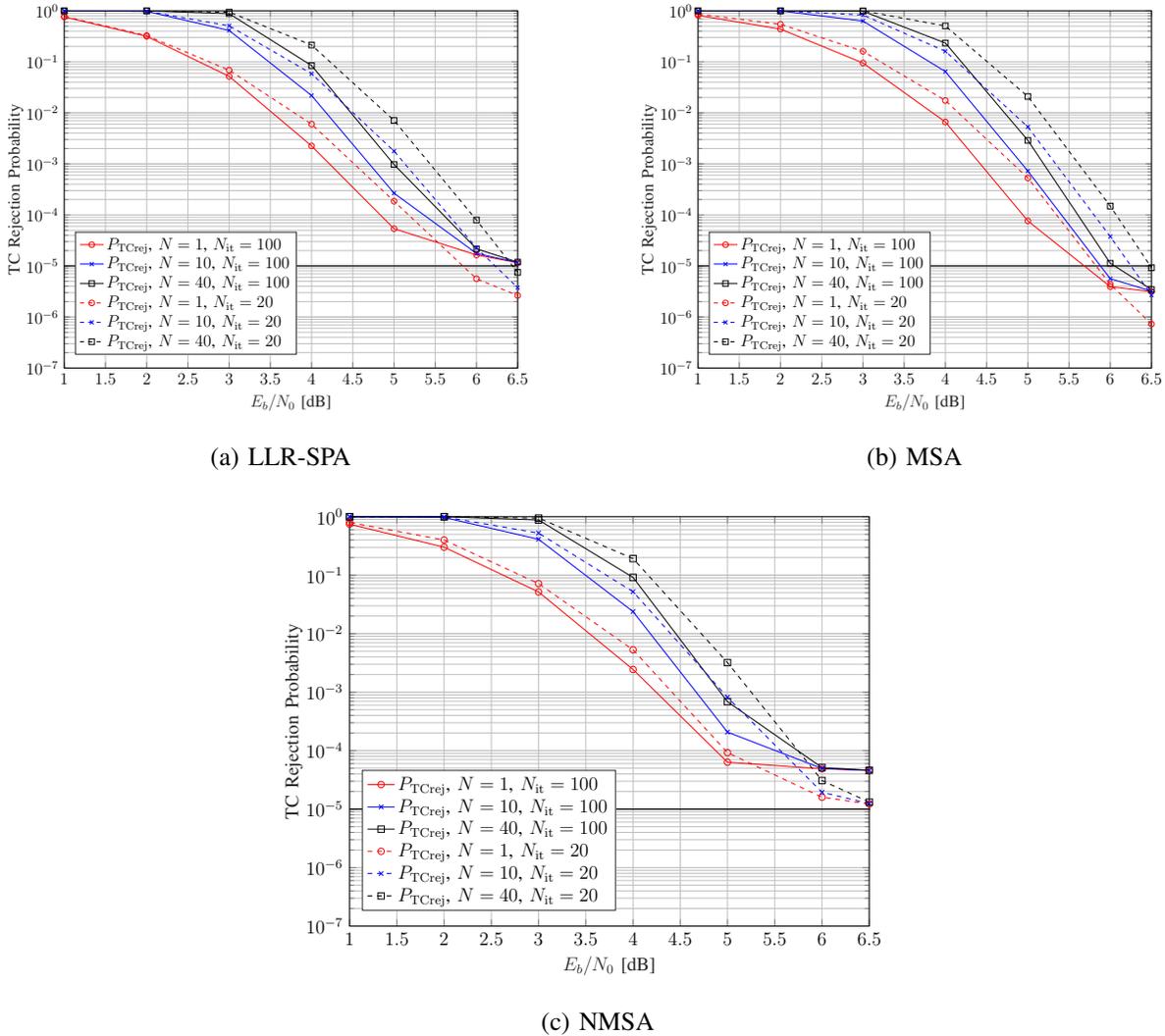
\begin{figure}
    \centering
    \begin{subfigure}[b]{0.48\textwidth}
        \centering
        \resizebox{\textwidth}{!}{\definecolor{mycolor1}{rgb}{1.00000,0.00000,1.00000}%
\definecolor{mycolor2}{rgb}{0.00000,1.00000,1.00000}%
\begin{tikzpicture}

\begin{axis}[%
width=4.521in,
height=3.566in,
at={(0.758in,0.481in)},
scale only axis,
unbounded coords=jump,
xmin=1,
xmax=6.5,
xlabel style={font=\color{white!15!black}},
xlabel={$E_b/N_0$ [dB]},
ymode=log,
ymin=1e-07,
ymax=1,
yminorticks=true,
ylabel style={font=\color{white!15!black}},
ylabel={TC Rejection Probability},
axis background/.style={fill=white},
title style={font=\bfseries},
xmajorgrids,
ymajorgrids,
yminorgrids,
legend style={at={(0.024,0.038)}, anchor=south west, legend cell align=left, align=left, draw=white!15!black}
]

    \draw[black, thick] (axis cs:1,1e-5) -- (axis cs:7,1e-5);

\addplot [color=red, mark=o, mark options={solid, red}]
  table[row sep=crcr]{%
0	1\\
        1	0.758006583890304\\
        2	0.31602817293063\\
        3	0.051775099964566\\
        4	0.002247768005941\\
        5	5.3699242441289e-05\\
        6	1.6414e-05\\
        6.5	1.16e-05\\
};
\addlegendentry{$P_{\mathrm{TCrej}}$, $N=1$, $N_{\mathrm{it}}=100$}

\addplot [color=blue, mark=x, mark options={solid, blue}]
  table[row sep=crcr]{%
0	1\\
1	0.999999311122747\\
2	0.9774773642425\\
3	0.411920897667465\\
4	0.0218378616925548\\
5	0.0002667363583115\\
6	1.7620e-05\\
6.5	1.1669e-05\\
};
\addlegendentry{$P_{\mathrm{TCrej}}$, $N=10$, $N_{\mathrm{it}}=100$}

\addplot [color=black, mark=square, mark options={solid, black}]
  table[row sep=crcr]{%
0	1\\
1	1\\
2	0.999999747520466\\
3	0.880378770032495\\
4	0.0843788062242400\\
5	0.0009736412579659\\
6	2.164e-05\\
6.5	1.19e-05\\
};
\addlegendentry{$P_{\mathrm{TCrej}}$, $N=40$, $N_{\mathrm{it}}=100$}

\addplot [color=red, dashed, mark=o, mark options={solid, red}]
  table[row sep=crcr]{%
0	1\\
  1	0.775004208905707\\
        2	0.327018567057995\\
        3	0.0685191890635125\\
        4	0.00600779277905918\\
        5	0.00018569\\
        6	5.603e-06\\
        6.5	2.6341e-06\\
};
\addlegendentry{$P_{\mathrm{TCrej}}$, $N=1$, $N_{\mathrm{it}}=20$}

\addplot [color=blue, dashed,  mark=x, mark options={solid, blue}]
  table[row sep=crcr]{%
0	1\\
1	0.999999667480547\\
2	0.980939308137935\\
3	0.508164756640705\\
4	0.0583277739339885\\
5	0.00177682626934843\\
6	2.2613e-05\\
6.5	3.755e-06\\
};
\addlegendentry{$P_{\mathrm{TCrej}}$, $N=10$, $N_{\mathrm{it}}=20$}

\addplot [color=black, dashed, mark=square, mark options={solid, black}]
  table[row sep=crcr]{%
0	1\\
1	1\\
2	0.999999867994937\\
3	0.941479851076784\\
4	0.213638273605542\\
5	0.0071\\
6	7.931e-05\\
6.5	7.4935e-06\\
};
\addlegendentry{$P_{\mathrm{TCrej}}$, $N=40$,  $N_{\mathrm{it}}=20$}

\end{axis}

\begin{axis}[%
width=5.833in,
height=4.375in,
at={(0in,0in)},
scale only axis,
xmin=0,
xmax=1,
ymin=0,
ymax=1,
axis line style={draw=none},
ticks=none,
axis x line*=bottom,
axis y line*=left
]
\end{axis}
\end{tikzpicture}%
}
        \caption{\ac{LLR-SPA}}
        \label{fig:spallr_randN}
    \end{subfigure}
    \hfill
    \begin{subfigure}[b]{0.48\textwidth}
        \centering
        \resizebox{\textwidth}{!}{\definecolor{mycolor1}{rgb}{1.00000,0.00000,1.00000}%
\definecolor{mycolor2}{rgb}{0.00000,1.00000,1.00000}%
\begin{tikzpicture}

\begin{axis}[%
width=4.521in,
height=3.566in,
at={(0.758in,0.481in)},
scale only axis,
unbounded coords=jump,
xmin=1,
xmax=6.5,
xlabel style={font=\color{white!15!black}},
xlabel={$E_b/N_0$ [dB]},
ymode=log,
ymin=1e-07,
ymax=1,
yminorticks=true,
ylabel style={font=\color{white!15!black}},
ylabel={TC Rejection Probability},
axis background/.style={fill=white},
title style={font=\bfseries},
xmajorgrids,
ymajorgrids,
yminorgrids,
legend style={at={(0.024,0.038)}, anchor=south west, legend cell align=left, align=left, draw=white!15!black}
]

    \draw[black, thick] (axis cs:1,1e-5) -- (axis cs:7,1e-5);

\addplot [color=red, mark=o, mark options={solid, red}]
  table[row sep=crcr]{%
       0	1\\
        1	0.813002649097218\\
        2	0.4384\\
        3	0.0948\\
        4	0.00661405303125915\\
        5	7.6307e-05\\
        6	3.9247e-06\\
        6.5	3.1314e-06\\
};
\addlegendentry{$P_{\mathrm{TCrej}}$, $N=1$, $N_{\mathrm{it}}=100$}

\addplot [color=blue, mark=x, mark options={solid, blue}]
  table[row sep=crcr]{%
0	1\\
1	0.999999947711051\\
2	0.996912357743470\\
3	0.6305\\
4	0.0644\\
5	0.000725512832915359\\
6	5.6257e-06\\
6.5	3.2054e-06\\
};
\addlegendentry{$P_{\mathrm{TCrej}}$, $N=10$, $N_{\mathrm{it}}=100$}

\addplot [color=black, mark=square, mark options={solid, black}]
  table[row sep=crcr]{%
0	1\\
1	1\\
2	0.999999999909111\\
3	0.9814\\
4	0.2337\\
5	0.00288757717445164\\
6	1.1296e-05\\
6.5	3.4520e-06\\
};
\addlegendentry{$P_{\mathrm{TCrej}}$, $N=40$, $N_{\mathrm{it}}=100$}

\addplot [color=red, dashed, mark=o, mark options={solid, red}]
  table[row sep=crcr]{%
 0	1\\
        1	0.862001564412852\\
        2	0.545000531750865\\
        3	0.161000747643103\\
        4	0.0174009285570585\\
        5	0.0005292602525926289\\
        6	4.3646e-06\\
        6.5	7.2710E-07\\
};
\addlegendentry{$P_{\mathrm{TCrej}}$, $N=1$, $N_{\mathrm{it}}=20$}

\addplot [color=blue, dashed,  mark=x, mark options={solid, blue}]
  table[row sep=crcr]{%
0	1\\
1	0.999999997495126\\
2	0.999584925214405\\
3	0.827218903020829\\
4	0.160989902541909\\
5	0.0053\\
6	3.7664e-05\\
6.5	2.6891e-06\\
};
\addlegendentry{$P_{\mathrm{TCrej}}$, $N=10$, $N_{\mathrm{it}}=20$}

\addplot [color=black, dashed, mark=square, mark options={solid, black}]
  table[row sep=crcr]{%
0	1\\
1	1\\
2	0.999999999999970\\
3	0.999149331487394\\
4	0.5043\\
5	0.0209\\
6	0.00014865\\
6.5	9.2291e-06\\
};
\addlegendentry{$P_{\mathrm{TCrej}}$, $N=40$, $N_{\mathrm{it}}=20$}

\end{axis}

\begin{axis}[%
width=5.833in,
height=4.375in,
at={(0in,0in)},
scale only axis,
xmin=0,
xmax=1,
ymin=0,
ymax=1,
axis line style={draw=none},
ticks=none,
axis x line*=bottom,
axis y line*=left
]
\end{axis}
\end{tikzpicture}%
}
        \caption{\ac{MSA}}
        \label{fig:ms_randN}
    \end{subfigure}
    
    \begin{subfigure}[b]{0.55\textwidth}
        \centering
        \resizebox{\textwidth}{!}{\definecolor{mycolor1}{rgb}{1.00000,0.00000,1.00000}%
\definecolor{mycolor2}{rgb}{0.00000,1.00000,1.00000}%
\begin{tikzpicture}

\begin{axis}[%
width=4.521in,
height=3.566in,
at={(0.758in,0.481in)},
scale only axis,
unbounded coords=jump,
xmin=1,
xmax=6.5,
xlabel style={font=\color{white!15!black}},
xlabel={$E_b/N_0$ [dB]},
ymode=log,
ymin=1e-07,
ymax=1,
yminorticks=true,
ylabel style={font=\color{white!15!black}},
ylabel={TC Rejection Probability},
axis background/.style={fill=white},
title style={font=\bfseries},
xmajorgrids,
ymajorgrids,
yminorgrids,
legend style={at={(0.024,0.038)}, anchor=south west, legend cell align=left, align=left, draw=white!15!black}
]

    \draw[black, thick] (axis cs:1,1e-5) -- (axis cs:7,1e-5);
    
\addplot [color=red, mark=o, mark options={solid, red}]
  table[row sep=crcr]{%
 0	1\\
        1	0.746015038278808\\
        2	0.302024561569932\\
        3	0.0517\\
        4	0.0024433956000594\\
        5	6.3329e-05\\
        6	4.896e-05\\
        6.5	4.5968e-05\\
};
\addlegendentry{$P_{\mathrm{TCrej}}$, $N=1$, $N_{\mathrm{it}}=100$}

\addplot [color=blue, mark=x, mark options={solid, blue}]
  table[row sep=crcr]{%
0	1\\
1	0.999998882335511\\
2	0.972550215215001\\
3	0.4119\\
4	0.023849191361283\\
5	0.00020731\\
6	4.9499e-05\\
6.5	4.5995e-05\\
};
\addlegendentry{$P_{\mathrm{TCrej}}$, $N=10$, $N_{\mathrm{it}}=100$}

\addplot [color=black, mark=square, mark options={solid, black}]
  table[row sep=crcr]{%
0	1\\
1	1\\
2	0.999999432191849\\
3	0.8804\\
4	0.0916803143140026\\
5	0.0006871\\
6	5.1297e-05\\
6.5	4.6084e-05\\
};
\addlegendentry{$P_{\mathrm{TCrej}}$, $N=40$, $N_{\mathrm{it}}=100$}

\addplot [color=red, dashed, mark=o, mark options={solid, red}]
  table[row sep=crcr]{%
        0	1\\
        1	0.800003921247626\\
        2	0.4000059990987\\
        3	0.0719\\
        4	0.0053\\
        5	9.2499025951289e-05\\
        6	1.598e-05\\
        6.5	1.2214e-05\\
};
\addlegendentry{$P_{\mathrm{TCrej}}$, $N=1$, $N_{\mathrm{it}}=20$}

\addplot [color=blue, dashed,  mark=x, mark options={solid, blue}]
  table[row sep=crcr]{%
0	1\\
1	0.999999897602008\\
2	0.993953442857093\\
3	0.5255\\
4	0.0521\\
5	0.0008167\\
6	1.94e-05\\
6.5	1.2434e-05\\
};
\addlegendentry{$P_{\mathrm{TCrej}}$, $N=10$, $N_{\mathrm{it}}=20$}

\addplot [color=black, dashed, mark=square, mark options={solid, black}]
  table[row sep=crcr]{%
0	1\\
1	1\\
2	0.999999998663264\\
3	0.9493\\
4	0.1925\\
5	0.0032\\
6	3.08e-05\\
6.5	1.3164e-05\\
};
\addlegendentry{$P_{\mathrm{TCrej}}$, $N=40$, $N_{\mathrm{it}}=20$}

\end{axis}

\begin{axis}[%
width=5.833in,
height=4.375in,
at={(0in,0in)},
scale only axis,
xmin=0,
xmax=1,
ymin=0,
ymax=1,
axis line style={draw=none},
ticks=none,
axis x line*=bottom,
axis y line*=left
]
\end{axis}
\end{tikzpicture}%
}
        \caption{\ac{NMSA}}
        \label{fig:nms_randN}
    \end{subfigure}
    
    \caption{TC rejection probability using different decoders and \ac{CLTU} lengths in the randomized case}
    \label{fig:combined_Ns_rand}
\end{figure}

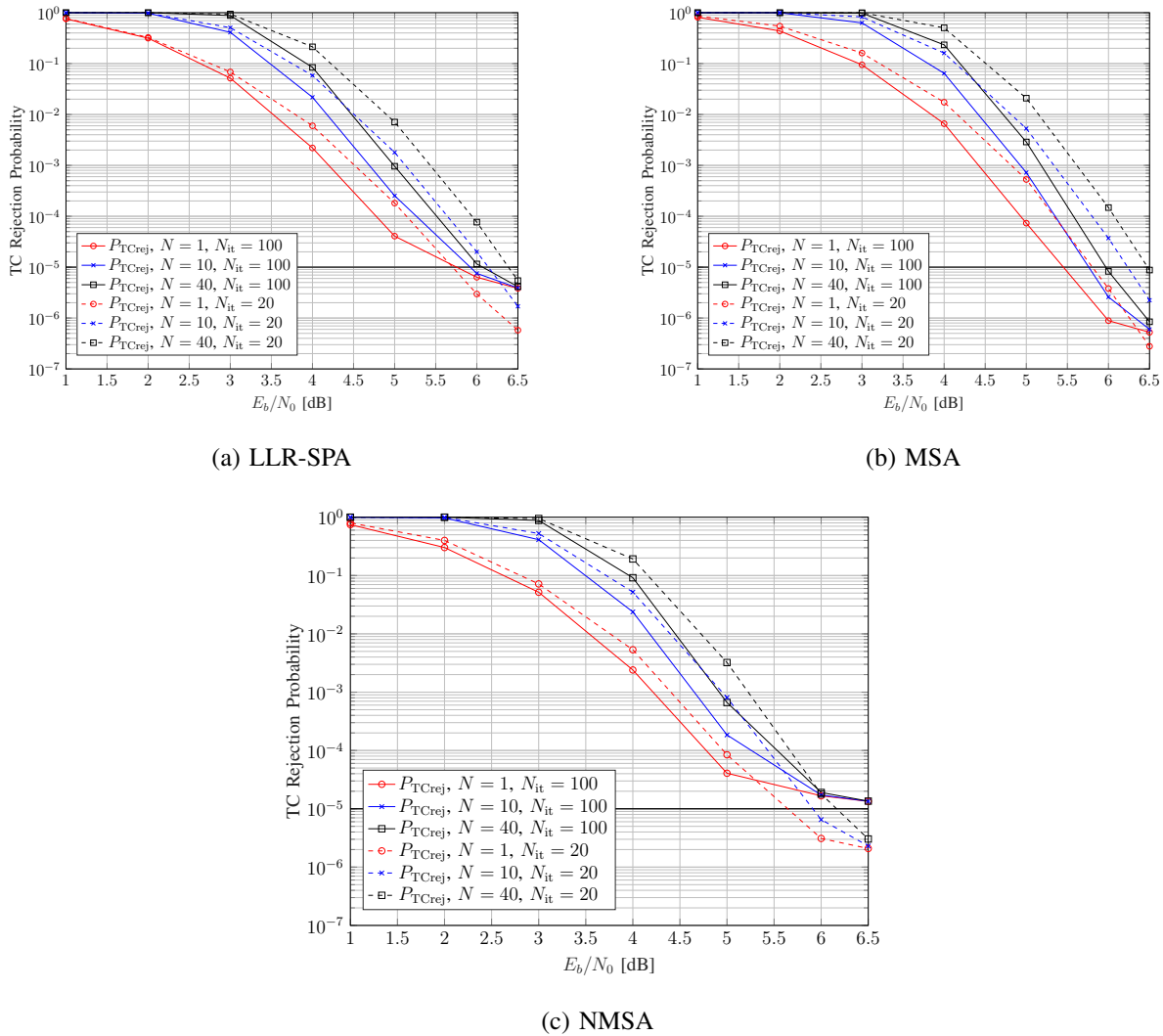
\begin{figure}
    \centering
    \begin{subfigure}[b]{0.48\textwidth}
        \centering
        \resizebox{\textwidth}{!}{\definecolor{mycolor1}{rgb}{1.00000,0.00000,1.00000}%
\definecolor{mycolor2}{rgb}{0.00000,1.00000,1.00000}%
\begin{tikzpicture}

\begin{axis}[%
width=4.521in,
height=3.566in,
at={(0.758in,0.481in)},
scale only axis,
unbounded coords=jump,
xmin=1,
xmax=6.5,
xlabel style={font=\color{white!15!black}},
xlabel={$E_b/N_0$ [dB]},
ymode=log,
ymin=1e-07,
ymax=1,
yminorticks=true,
ylabel style={font=\color{white!15!black}},
ylabel={TC Rejection Probability},
axis background/.style={fill=white},
title style={font=\bfseries},
xmajorgrids,
ymajorgrids,
yminorgrids,
legend style={at={(0.024,0.038)}, anchor=south west, legend cell align=left, align=left, draw=white!15!black}
]

    \draw[black, thick] (axis cs:1,1e-5) -- (axis cs:7,1e-5);

\addplot [color=red, mark=o, mark options={solid, red}]
  table[row sep=crcr]{%
0	1\\
        1	0.758\\
        2	0.3155\\ 
        3	0.0517\\ 
        4	0.0022\\ 
        5	4.07E-05\\ 
        6	6.3320E-06\\ 
        6.5 3.8352E-06\\
};
\addlegendentry{$P_{\mathrm{TCrej}}$, $N=1$, $N_{\mathrm{it}}=100$}

\addplot [color=blue, mark=x, mark options={solid, blue}]
  table[row sep=crcr]{%
0	1\\
1	0.999999311118201\\
2	0.977584475507661\\
3	0.411918721665769\\
4	0.0218145801392087\\
5	0.00025308\\
6	7.538e-06\\
6.5 3.9044E-06\\
};
\addlegendentry{$P_{\mathrm{TCrej}}$, $N=10$, $N_{\mathrm{it}}=100$}

\addplot [color=black, mark=square, mark options={solid, black}]
  table[row sep=crcr]{%
0	1\\
1	1\\
2	0.999999747517108\\
3	0.880378327411769\\
4	0.0843570132281376\\
5	0.00096066\\
6	1.1558e-05\\
6.5 4.1351E-06\\
};
\addlegendentry{$P_{\mathrm{TCrej}}$, $N=40$, $N_{\mathrm{it}}=100$}

\addplot [color=red, dashed, mark=o, mark options={solid, red}]
  table[row sep=crcr]{%
0	1\\
        1	0.775003500163153\\
        2	0.327010894858674\\
        3	0.0685192822135125\\
        4	0.00600093411005918\\
        5	0.0001812009113231289\\
        6	2.97e-06\\
        6.5 5.78E-07\\
};
\addlegendentry{$P_{\mathrm{TCrej}}$, $N=1$, $N_{\mathrm{it}}=20$}

\addplot [color=blue, dashed,  mark=x, mark options={solid, blue}]
  table[row sep=crcr]{%
0	1\\
1	0.999999667479500\\
2	0.980939090840072\\
3	0.508164805825243\\
4	0.0583212762793206\\
5	0.0018\\
6	1.998239417010e-05\\
6.5 1.6994E-06\\
};
\addlegendentry{$P_{\mathrm{TCrej}}$, $N=10$, $N_{\mathrm{it}}=20$}

\addplot [color=black, dashed, mark=square, mark options={solid, black}]
  table[row sep=crcr]{%
0	1\\
1	1\\
2	0.999999867993432\\
3	0.941479856928920\\
4	0.2135\\
5	0.0071\\
6	7.6677e-05\\
6.5 5.4374E-06\\
};
\addlegendentry{$P_{\mathrm{TCrej}}$, $N=40$,  $N_{\mathrm{it}}=20$}

\end{axis}

\begin{axis}[%
width=5.833in,
height=4.375in,
at={(0in,0in)},
scale only axis,
xmin=0,
xmax=1,
ymin=0,
ymax=1,
axis line style={draw=none},
ticks=none,
axis x line*=bottom,
axis y line*=left
]
\end{axis}
\end{tikzpicture}%
}
        \caption{\ac{LLR-SPA}}
        \label{fig:N_spallr_notrand}
    \end{subfigure}
    \hfill
    \begin{subfigure}[b]{0.48\textwidth}
        \centering
        \resizebox{\textwidth}{!}{\definecolor{mycolor1}{rgb}{1.00000,0.00000,1.00000}%
\definecolor{mycolor2}{rgb}{0.00000,1.00000,1.00000}%
\begin{tikzpicture}

\begin{axis}[%
width=4.521in,
height=3.566in,
at={(0.758in,0.481in)},
scale only axis,
unbounded coords=jump,
xmin=1,
xmax=6.5,
xlabel style={font=\color{white!15!black}},
xlabel={$E_b/N_0$ [dB]},
ymode=log,
ymin=1e-07,
ymax=1,
yminorticks=true,
ylabel style={font=\color{white!15!black}},
ylabel={TC Rejection Probability},
axis background/.style={fill=white},
title style={font=\bfseries},
xmajorgrids,
ymajorgrids,
yminorgrids,
legend style={at={(0.024,0.038)}, anchor=south west, legend cell align=left, align=left, draw=white!15!black}
]

    \draw[black, thick] (axis cs:1,1e-5) -- (axis cs:7,1e-5);

\addplot [color=red, mark=o, mark options={solid, red}]
  table[row sep=crcr]{%
0	1\\
        1	0.813002456489242\\
        2	0.438400\\
        3	0.0948\\
        4	0.0066\\
        5	7.3213e-05\\
        6	8.89e-07\\
        6.5 5.2472E-07\\
};
\addlegendentry{$P_{\mathrm{TCrej}}$, $N=1$, $N_{\mathrm{it}}=100$}

\addplot [color=blue, mark=x, mark options={solid, blue}]
  table[row sep=crcr]{%
0	1\\
1	0.999999947710997\\
2	0.996912356292273\\
3	0.6305\\
4	0.0644\\
5	0.00072242\\
6	2.59e-06\\
6.5 5.987E-07\\
};
\addlegendentry{$P_{\mathrm{TCrej}}$, $N=10$, $N_{\mathrm{it}}=100$}

\addplot [color=black, mark=square, mark options={solid, black}]
  table[row sep=crcr]{%
0	1\\
1	1\\
2	0.999999999909111\\
3	0.9814\\
4	0.2337\\
5	0.00288498467268966\\
6	8.26396681649340e-06\\
6.5 8.453E-07\\
};
\addlegendentry{$P_{\mathrm{TCrej}}$, $N=40$, $N_{\mathrm{it}}=100$}

\addplot [color=red, dashed, mark=o, mark options={solid, red}]
  table[row sep=crcr]{%
        0	1\\
        1	0.862001542609082\\
        2	0.5450\\
        3	0.161000446801103\\
        4	0.0174003350666585\\
        5	0.00052861\\
        6	3.8043e-06\\
        6.5 2.7855E-07\\
};
\addlegendentry{$P_{\mathrm{TCrej}}$, $N=1$, $N_{\mathrm{it}}=20$}

\addplot [color=blue, dashed,  mark=x, mark options={solid, blue}]
  table[row sep=crcr]{%
0	1\\
1	0.999999997495125\\
2	0.999584925020565\\
3	0.8272\\
4	0.16091\\
5	0.0053\\
6	3.7104e-05\\
6.5 2.2405E-06\\
};
\addlegendentry{$P_{\mathrm{TCrej}}$, $N=10$, $N_{\mathrm{it}}=20$}

\addplot [color=black, dashed, mark=square, mark options={solid, black}]
  table[row sep=crcr]{%
0	1\\
1	1\\
2	0.999999999999970\\
3	0.999149331182003\\
4	0.5043\\
5	0.0209\\
6	0.000148091\\
6.5 8.7805E-06\\
};
\addlegendentry{$P_{\mathrm{TCrej}}$, $N=40$, $N_{\mathrm{it}}=20$}

\end{axis}

\begin{axis}[%
width=5.833in,
height=4.375in,
at={(0in,0in)},
scale only axis,
xmin=0,
xmax=1,
ymin=0,
ymax=1,
axis line style={draw=none},
ticks=none,
axis x line*=bottom,
axis y line*=left
]
\end{axis}
\end{tikzpicture}%
}
        \caption{\ac{MSA}}
        \label{fig:N_ms_notrand}
    \end{subfigure}
    
    \begin{subfigure}[b]{0.55\textwidth}
        \centering
        \resizebox{\textwidth}{!}{\definecolor{mycolor1}{rgb}{1.00000,0.00000,1.00000}%
\definecolor{mycolor2}{rgb}{0.00000,1.00000,1.00000}%
\begin{tikzpicture}

\begin{axis}[%
width=4.521in,
height=3.566in,
at={(0.758in,0.481in)},
scale only axis,
unbounded coords=jump,
xmin=1,
xmax=6.5,
xlabel style={font=\color{white!15!black}},
xlabel={$E_b/N_0$ [dB]},
ymode=log,
ymin=1e-07,
ymax=1,
yminorticks=true,
ylabel style={font=\color{white!15!black}},
ylabel={TC Rejection Probability},
axis background/.style={fill=white},
title style={font=\bfseries},
xmajorgrids,
ymajorgrids,
yminorgrids,
legend style={at={(0.024,0.038)}, anchor=south west, legend cell align=left, align=left, draw=white!15!black}
]

    \draw[black, thick] (axis cs:1,1e-5) -- (axis cs:7,1e-5);
    
\addplot [color=red, mark=o, mark options={solid, red}]
  table[row sep=crcr]{%
0	1\\
        1	0.746011177519371\\
        2	0.302020722570272\\
        3	0.0517\\
        4	0.0024\\
        5	4.069e-05\\
        6	1.67e-05\\
        6.5 1.3418E-05\\
};
\addlegendentry{$P_{\mathrm{TCrej}}$, $N=1$, $N_{\mathrm{it}}=100$}

\addplot [color=blue, mark=x, mark options={solid, blue}]
  table[row sep=crcr]{%
0	1\\
1	0.999998882318521\\
2	0.972550064235885\\
3	0.412\\
4	0.0237667607457343\\
5	0.00018467\\
6	1.7239e-05\\
6.5 1.3445E-05\\
};
\addlegendentry{$P_{\mathrm{TCrej}}$, $N=10$, $N_{\mathrm{it}}=100$}

\addplot [color=black, mark=square, mark options={solid, black}]
  table[row sep=crcr]{%
0	1\\
1	1\\
2	0.999999432188726\\
3	0.8804\\
4	0.0916634188328954\\
5	0.00066447\\
6	1.9036e-05\\
6.5 1.3534E-05\\
};
\addlegendentry{$P_{\mathrm{TCrej}}$, $N=40$, $N_{\mathrm{it}}=100$}

\addplot [color=red, dashed, mark=o, mark options={solid, red}]
  table[row sep=crcr]{%
        0	1\\
        1	0.800003557251450\\
        2	0.400004913098796\\
        3	0.0718061912569355\\
        4	0.00533642556825922\\
        5	8.443e-05\\
        6	3.1e-06\\
        6.5 2.0844E-06\\
};
\addlegendentry{$P_{\mathrm{TCrej}}$, $N=1$, $N_{\mathrm{it}}=20$}

\addplot [color=blue, dashed,  mark=x, mark options={solid, blue}]
  table[row sep=crcr]{%
0	1\\
1	0.999999897601821\\
2	0.993953431912716\\
3	0.5255\\
4	0.0520\\
5	0.00080864\\
6	6.52e-06\\
6.5 2.3036E-06\\
};
\addlegendentry{$P_{\mathrm{TCrej}}$, $N=10$, $N_{\mathrm{it}}=20$}

\addplot [color=black, dashed, mark=square, mark options={solid, black}]
  table[row sep=crcr]{%
0	1\\
1	1\\
2	0.999999998663262\\
3	0.9493\\
4	0.192471285731548\\
5	0.00321891775966604\\
6	1.79198454133385e-05\\
6.5 3.0344E-06\\
};
\addlegendentry{$P_{\mathrm{TCrej}}$, $N=40$, $N_{\mathrm{it}}=20$}

\end{axis}

\begin{axis}[%
width=5.833in,
height=4.375in,
at={(0in,0in)},
scale only axis,
xmin=0,
xmax=1,
ymin=0,
ymax=1,
axis line style={draw=none},
ticks=none,
axis x line*=bottom,
axis y line*=left
]
\end{axis}
\end{tikzpicture}%
}
        \caption{\ac{NMSA}}
        \label{fig:N_nms_notrand}
    \end{subfigure}
    
    \caption{TC Rejection Probability for different decoders and \ac{CLTU} lengths, when using the proposed \ac{TS} in the de-randomized case}
    \label{fig:combined_Ns_notrand}
\end{figure}

In this section we extend the analysis to the case of $N > 1$, that is, the \ac{CLTU} contains more than one codeword. This reflects a more general scenario since, in space missions, data longer than one word clearly span over more codewords. As an example, we compare the performance of the system when transmitting a \ac{CLTU} with encoded data consisting of a single codeword ($N = 1$) to those with encoded data consisting of $N = 10$ and $N = 40$ codewords. As expected from the discussion in Section \ref{sec:numres}-\ref{subsec:randotc}, the system performance is worse in the case of a \ac{CLTU} containing $10$ and $40$ codewords than in the case with a single codeword. This is clearly shown in Fig.s \ref{fig:combined_Ns_rand} and \ref{fig:combined_Ns_notrand} for the randomized and the de-randomized case, respectively. 
According to \eqref{eq:cerN}, such a result would also be confirmed by considering other values of $N>1$.
So, the case with $N = 1$ should be seen as the best scenario. Notice that, generalizing the analysis, we have kept the same system requirements as in Section \ref{sec:numres}-\ref{subsec:randotc}.

From Fig.s \ref{fig:combined_Ns_rand} and \ref{fig:combined_Ns_notrand}, we also observe that, when $E_b/N_0$ is small (say not larger than $5$ dB), in both the randomized  and the de-randomized case decreasing the number of iterations does not yield any advantage in terms of \ac{TC} rejection probability, independently of $N$. This perfectly aligns with previous Fig.s \ref{fig:tcrejran} and \ref{fig:tcrej_nr} (characterized by $N=1$), where the reduction of the maximum number of iterations also leads to a reduction in the \ac{TC} rejection probability when $E_b/N_0$ is larger than $5$. However, in the (relatively) large $E_b/N_0$ regime, this might not happen when $N=10$ or $N=40$ (corresponding to the blue and black curves in Fig.s \ref{fig:combined_Ns_rand} and \ref{fig:combined_Ns_notrand}, respectively). For example, when employing the \ac{MSA}-based decoder, with $N=10$ or $N=40$, decreasing the maximum number of iterations from $100$ to $20$ does not yield any advantage for any of the considered values of $E_b/N_0$, especially when $N=10$. On the contrary, when the \ac{NMSA}-based decoder is used, for relatively high values of $E_b/N_0$, the system performance benefits from the reduction of the maximum number of decoding iterations. This is a consequence of the fact that the \ac{NMSA} is the decoding algorithm with the smallest \ac{CER}. Therefore, since $N$ and $E_b/N_0$ are both not very large, the leading term in \eqref{eq:P_TCrej} remains the not-acknowledged termination probability $P_{\mathrm{nat}}$. 

Generally, Fig.s \ref{fig:combined_Ns_rand} and \ref{fig:combined_Ns_notrand} illustrate that, depending on the considered decoder, the operating point and the \ac{CLTU} length, and assuming the \ac{TS} is employed and detected using a decoder, it may be necessary to optimize the number of iterations used. It is indeed evident that increasing the maximum number of iterations reduces the \ac{CER} but increases the $P_{\mathrm{nat}}$, as the additional iterations aid decoder convergence. In other words, the increase of the  maximum number of iterations is beneficial for coded data, since it helps the decoder to converge more often to a codeword, but undesirable for the \ac{TS}, which should instead trigger a decoding error. Therefore, finding the optimal trade-off might be crucial.

\subsection{Remarks on large values of the signal-to-noise ratio}

Running Monte Carlo simulations to estimate \ac{CER} values on the order of $10^{-8}$ or lower is computationally demanding. For the considered $(128,64)$ \ac{LDPC} code, this significantly hinders numerical analysis for $E_b/N_0$ values of $7$ dB or higher. However, assuming that the \ac{CER} continues to drop faster than $P_\mathrm{nat}$ for values slightly greater than $7$ dB, we can leverage the insights from Section \ref{sec:tcrej}-\ref{subsec:LDPCdecodefail} to infer some conclusions. In particular, according to \eqref{eq:approximationCER}, for small to moderate values of $N$, we can expect that the leading term in \eqref{eq:P_TCrej} is the not-acknowledged termination probability $P_{\mathrm{nat}}$. In other words, we expect the trend of the \ac{TC} rejection probability to be determined by the not-acknowledged termination probability. However, we also remark that, for extremely large values of $E_b/N_0$, the \ac{TS} (especially in the de-randomized case) always causes a decoding error, eventually yielding $P_{\mathrm{nat}}\rightarrow 0$ when $E_b/N_0 \rightarrow \infty$. Therefore, clearly, if the error rate performance of the \ac{LDPC} code exhibits an error floor at any point, a more comprehensive analysis will be necessary. This entails delving into the causes of the error floor, which may include investigating specific code properties and decoding algorithm limitations, such as trapping sets \cite{Richardson2003a}, pseudo-codewords \cite{Koetter2003}, and other harmful objects \cite{Battaglioni2023Mitch}. This analysis goes beyond the scope of this paper and is left for future works.

\section{Conclusion and future works}\label{sec:concl}

We have analyzed the performance of a \ac{CCSDS}-compliant communication scheme incorporating \ac{LDPC}-coded transmissions in satellite systems, with a focus on the optional tail sequence that may be required for termination. We have evaluated the \ac{TC} rejection probability, both theoretically and numerically, across various operation conditions, involving the length of the \ac{CLTU}, the decoding algorithm and the maximum number of decoding iterations. We have demonstrated that a well-designed tail sequence, which is sufficiently distant from any valid codeword according to a chosen metric, effectively triggers a decoding error, thereby aiding correct termination of the \ac{CLTU}.

As mentioned, we have evaluated the performance using different iterative decoding algorithms, namely the \ac{LLR-SPA}, the \ac{MSA}, and the \ac{NMSA}. Our results show that, in the considered setting, the \ac{MSA}-based decoder performs better than the other two algorithms when the optional \ac{TS} is employed to trigger a decoding error. Our analysis also indicates that reducing the maximum number of decoding iterations does not always provide an advantage in terms of \ac{TC} rejection probability, particularly when $E_b/N_0$ is below $5$ dB. However, for extremely short \acp{CLTU} and for moderate to high values of $E_b/N_0$, reducing the number of iterations can enhance performance, from the \ac{TS} detection viewpoint, highlighting the importance of tailoring the decoding strategy to the specific context.

Through a combination of theoretical analysis and Monte Carlo simulations, we have provided a comprehensive understanding of the factors affecting \ac{TC} rejection probability.

As a catalyst for future research, we intend to explore a novel approach to the design of the \ac{TS}. Specifically, we plan to focus on harmful structures that induce decoding errors, such as trapping sets, absorbing sets, and fully absorbing sets. While the current method of designing a \ac{TS} that is distant from all codewords is based on the general \ac{ML} decoding principle, an alternative approach is to consider the unique characteristics of iterative decoders commonly used for decoding \ac{LDPC} codes. This is crucial because harmful patterns are not necessarily very distant from codewords.

\appendices

\section{$(128,64)$ and $(512,256)$ LDPC codes description} \label{app:LDPC12864}

The $(128,64)$ \ac{LDPC} code is specified by an $m\times n$ parity-check matrix $\mathbf{H}_{64 \times 128}$, where $m=n-k=128-64=64$ and $n=128$. This matrix is constructed from $M\times M$ submatrices, where $M=k/4=n/8=16$, as follows:

\[
\mathbf{H}_{64 \times 128} = \begin{bmatrix}
    \mathbf{I}_M \oplus \mathbf{\Phi}^7 & \mathbf{\Phi}^2 & \mathbf{\Phi}^{14} & \mathbf{\Phi}^6 & \mathbf{0}_M & \mathbf{\Phi}^0 & \mathbf{\Phi}^{13} & \mathbf{I}_M \\
    \mathbf{\Phi}^6 & \mathbf{I}_M \oplus \mathbf{\Phi}^{15} & \mathbf{\Phi}^0 & \mathbf{\Phi}^1 & \mathbf{I}_M & \mathbf{0}_M & \mathbf{\Phi}^0 & \mathbf{\Phi}^7 \\
    \mathbf{\Phi}^4 & \mathbf{\Phi}^1 & \mathbf{I}_M \oplus \mathbf{\Phi}^{15} & \mathbf{\Phi}^{14} & \mathbf{\Phi}^{11} & \mathbf{I}_M & \mathbf{0}_M & \mathbf{\Phi}^3 \\
    \mathbf{\Phi}^0 & \mathbf{\Phi}^1 & \mathbf{\Phi}^9 & \mathbf{I}_M \oplus \mathbf{\Phi}^{13} & \mathbf{\Phi}^{14} & \mathbf{\Phi}^1 & \mathbf{I}_M & \mathbf{0}_M \\
\end{bmatrix}, 
\]
where $\mathbf{I}_M$ is the $M \times M$ identity matrix, $\mathbf{\Phi}^i$ is the $i$-th right circular shift of $\mathbf{I}_M$, where ${0\leq i \leq M-1 }$, and $\mathbf{0}_M$ is the $M \times M$ zero matrix. Finally, the $\oplus$ operator indicates modulo-2 addition.

Notice that the structure of the $(512,256)$ \ac{LDPC} code's parity-check matrix is as follows:

\[ \mathbf{H} = \begin{bmatrix}
    \mathbf{I}_M \oplus \mathbf{\Phi}^{63} & \mathbf{\Phi}^{30} & \mathbf{\Phi}^{50} & \mathbf{\Phi}^{25} & \mathbf{0}_M & \mathbf{\Phi}^{43} & \mathbf{\Phi}^{62} & \mathbf{I}_M \\
    \mathbf{\Phi}^{56} & \mathbf{I}_M \oplus \mathbf{\Phi}^{61} & \mathbf{\Phi}^{50} & \mathbf{\Phi}^{25} & \mathbf{I}_M & \mathbf{0}_M & \mathbf{\Phi}^{37} & \mathbf{\Phi}^{26} \\
    \mathbf{\Phi}^{16} & \mathbf{\Phi}^0 & \mathbf{I}_M \oplus \mathbf{\Phi}^{55} & \mathbf{\Phi}^{27} & \mathbf{\Phi}^{56} & \mathbf{I}_M & \mathbf{0}_M & \mathbf{\Phi}^{43} \\
    \mathbf{\Phi}^{35} & \mathbf{\Phi}^{56} & \mathbf{\Phi}^{62} & \mathbf{I}_M \oplus \mathbf{\Phi}^{11} & \mathbf{\Phi}^{58} & \mathbf{\Phi}^3 & \mathbf{I}_M & \mathbf{0}_M \\ 
\end{bmatrix}, \] 
where $M = k/4, n/8 = 32$, is the same as the one of the $(128,64)$ \ac{LDPC} code. However, it is important to remark, as already highlighted in Section \ref{sec:preli}-\ref{subsec:stdcomsys}, that, when the $(512,256)$ \ac{LDPC} code is used, the tail sequence is not considered. 

\section{Decoding success rate for different values of $E_b/N_0$} \label{app:Hystograms}

In this appendix, we provide further analysis to support the results presented in Fig. \ref{fig:combined_pnats}, expanding the discussion in Section \ref{sec:numres}-\ref{subsec:decodingsuccesses}.  In particular, we present the results in Fig.s \ref{fig:combined_occ_rand} and \ref{fig:combined_occ_notrand}, in terms of decoding success rates, categorized by each considered value of \(E_b/N_0\). Let us remind that data were collected by analyzing $3\,000\,000$ transmissions of the randomized and de-randomized \ac{TS}, with $100$ maximum decoding iterations.

\begin{figure}
    \centering
    \begin{subfigure}[b]{0.48\textwidth}
        \centering
        \resizebox{\textwidth}{!}{
        \definecolor{mycolor1}{rgb}{0.00000,0.44700,0.74100}%
\begin{tikzpicture}

\begin{axis}[%
width=4.521in,
height=3.566in,
at={(0.758in,0.481in)},
scale only axis,
bar shift auto,
xmin=-0.2,
xmax=7.5,
xlabel style={font=\color{white!15!black}},
xlabel={$E_b$/$N_0$ [dB]},
ymin=0,
ymax=1,
ylabel style={font=\color{white!15!black}},
ylabel={Decoding success rate},
axis background/.style={fill=white},
grid=both, 
major grid style={line width=.2pt,draw=gray!50},
minor grid style={line width=.1pt,draw=gray!20}
]
\addplot[ybar, bar width=5, fill=mycolor1, draw=mycolor1, area legend] table[row sep=crcr] {%
0   0.0086\\
1	0.0648\\
2	0.1859\\
3	0.2334\\
4	0.1960\\
5	0.1643\\
6	0.0980\\
7	0.0490\\
};    
\addplot[forget plot, color=white!15!black] table[row sep=crcr] {%
-0.2	0\\
7.5	0\\
};
\end{axis}

\begin{axis}[%
width=5.833in,
height=4.375in,
at={(0in,0in)},
scale only axis,
xmin=0,
xmax=1,
ymin=0,
ymax=1,
axis line style={draw=none},
ticks=none,
axis x line*=bottom,
axis y line*=left
]
\end{axis}
\end{tikzpicture}%
        }
        \caption{\ac{LLR-SPA}}
        \label{fig:subspa}
    \end{subfigure}
    \hfill
    \begin{subfigure}[b]{0.48\textwidth}
        \centering
        \resizebox{\textwidth}{!}{
        \definecolor{mycolor1}{rgb}{0.00000,0.44700,0.74100}%
\begin{tikzpicture}

\begin{axis}[%
width=4.521in,
height=3.566in,
at={(0.758in,0.481in)},
scale only axis,
bar shift auto,
xmin=-0.2,
xmax=7.5,
xlabel style={font=\color{white!15!black}},
xlabel={$E_b$/$N_0$ [dB]},
ymin=0,
ymax=1,
ylabel style={font=\color{white!15!black}},
ylabel={Decoding success rate},
axis background/.style={fill=white},
grid=both, 
major grid style={line width=.2pt,draw=gray!50},
minor grid style={line width=.1pt,draw=gray!20}
]
\addplot[ybar, bar width=5, fill=mycolor1, draw=mycolor1, area legend] table[row sep=crcr] {%
0   0.1203\\
1	0.0902\\
2	0.0977\\
3	0.1504\\
4	0.1805\\
5	0.1429\\
6	0.1353\\
7	0.0827\\
};       
\addplot[forget plot, color=white!15!black] table[row sep=crcr] {%
-0.2	0\\
7.5	0\\
};
\end{axis}

\begin{axis}[%
width=5.833in,
height=4.375in,
at={(0in,0in)},
scale only axis,
xmin=0,
xmax=1,
ymin=0,
ymax=1,
axis line style={draw=none},
ticks=none,
axis x line*=bottom,
axis y line*=left
]
\end{axis}
\end{tikzpicture}%
        }
        \caption{\ac{MSA}}
    \end{subfigure}
    \hfill
    \begin{subfigure}[b]{0.48\textwidth}
        \centering
        \resizebox{\textwidth}{!}{
        \definecolor{mycolor1}{rgb}{0.00000,0.44700,0.74100}%
\begin{tikzpicture}

\begin{axis}[%
width=4.521in,
height=3.566in,
at={(0.758in,0.481in)},
scale only axis,
bar shift auto,
xmin=-0.2,
xmax=7.5,
xlabel style={font=\color{white!15!black}},
xlabel={$E_b$/$N_0$ [dB]},
ymin=0,
ymax=1,
ylabel style={font=\color{white!15!black}},
ylabel={Decoding success rate},
axis background/.style={fill=white},
grid=both, 
major grid style={line width=.2pt,draw=gray!50},
minor grid style={line width=.1pt,draw=gray!20}
]
\addplot[ybar, bar width=5, fill=mycolor1, draw=mycolor1, area legend] table[row sep=crcr] {%
0   0.1053\\
1	0.1142\\
2	0.1142\\
3	0.1133\\
4	0.1427\\  
5	0.1534\\
6	0.1427\\
7	0.1142\\
};   
\addplot[forget plot, color=white!15!black] table[row sep=crcr] {%
-0.2	0\\
7.5	0\\
};
\end{axis}

\begin{axis}[%
width=5.833in,
height=4.375in,
at={(0in,0in)},
scale only axis,
xmin=0,
xmax=1,
ymin=0,
ymax=1,
axis line style={draw=none},
ticks=none,
axis x line*=bottom,
axis y line*=left
]
\end{axis}
\end{tikzpicture}%
        }
        \caption{\ac{NMSA}}
    \end{subfigure}
    \caption{Decoding success rate for different $E_b$/$N_0$ values in the randomized case}
    \label{fig:ist_freq_rand}
\end{figure}
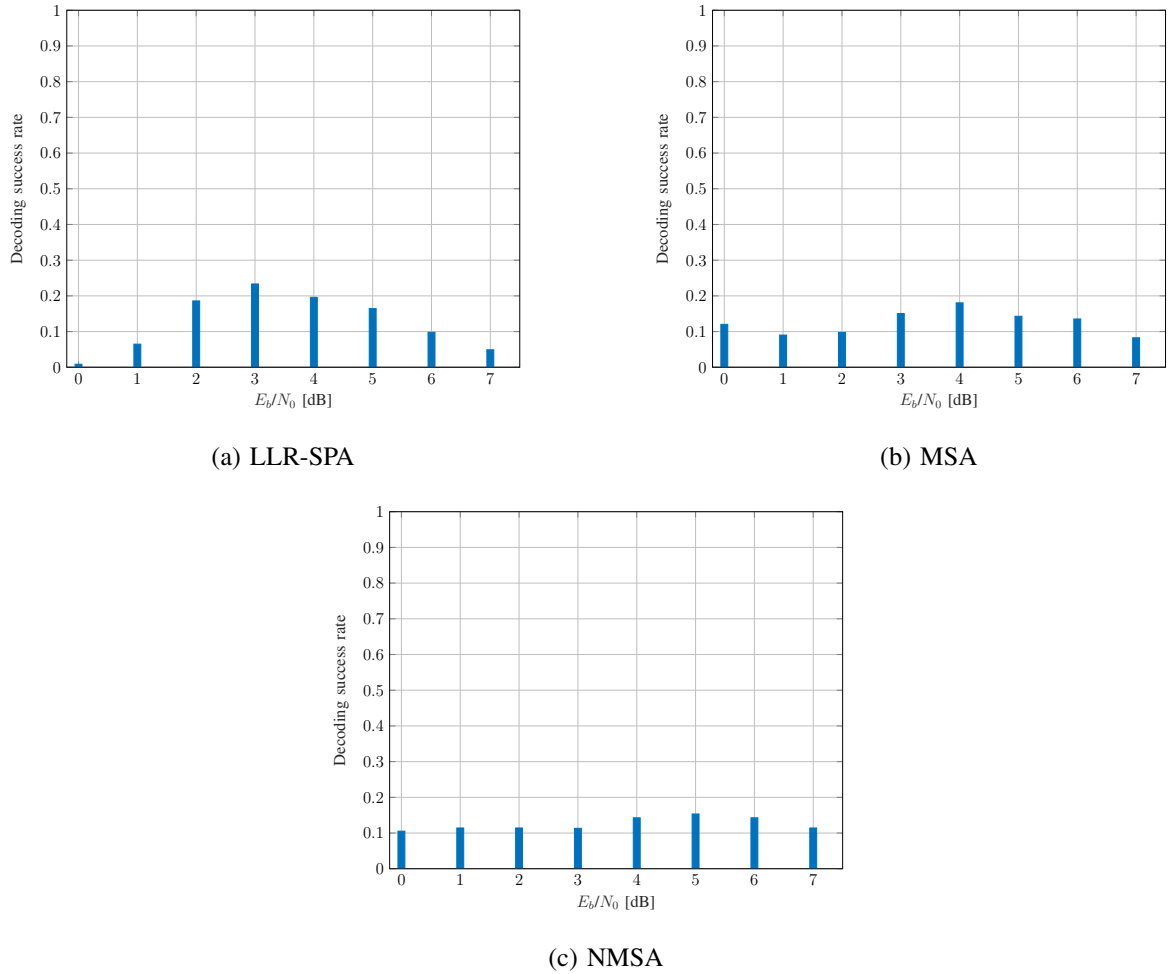

\begin{figure}
    \centering
    \begin{subfigure}[b]{0.48\textwidth}
        \centering
        \resizebox{\textwidth}{!}{
        \definecolor{mycolor1}{rgb}{0.00000,0.44700,0.74100}%
\begin{tikzpicture}

\begin{axis}[%
width=4.521in,
height=3.566in,
at={(0.758in,0.481in)},
scale only axis,
bar shift auto,
xmin=-0.2,
xmax=7.5,
xlabel style={font=\color{white!15!black}},
xlabel={$E_b$/$N_0$ [dB]},
ymin=0,
ymax=1,
ylabel style={font=\color{white!15!black}},
ylabel={Decoding success rate},
axis background/.style={fill=white},
grid=both, 
major grid style={line width=.2pt,draw=gray!50},
minor grid style={line width=.1pt,draw=gray!20}
]
\addplot[ybar, bar width=5, fill=mycolor1, draw=mycolor1, area legend] table[row sep=crcr] {%
0   0.0129\\
1	0.0860\\
2	0.2129\\
3	0.2796\\
4	0.2516\\
5	0.1204\\
6	0.0344\\
7	0.0022\\
};            
\addplot[forget plot, color=white!15!black] table[row sep=crcr] {%
-0.2	0\\
7.5	0\\
};
\end{axis}

\begin{axis}[%
width=5.833in,
height=4.375in,
at={(0in,0in)},
scale only axis,
xmin=0,
xmax=1,
ymin=0,
ymax=1,
axis line style={draw=none},
ticks=none,
axis x line*=bottom,
axis y line*=left
]
\end{axis}
\end{tikzpicture}%
        }
        \caption{\ac{LLR-SPA}}
        \label{fig:subspader}
    \end{subfigure}
    \hfill
    \begin{subfigure}[b]{0.48\textwidth}
        \centering
        \resizebox{\textwidth}{!}{
        \definecolor{mycolor1}{rgb}{0.00000,0.44700,0.74100}%
\begin{tikzpicture}

\begin{axis}[%
width=4.521in,
height=3.566in,
at={(0.758in,0.481in)},
scale only axis,
bar shift auto,
xmin=-0.2,
xmax=7.5,
xlabel style={font=\color{white!15!black}},
xlabel={$E_b$/$N_0$ [dB]},
ymin=0,
ymax=1,
ylabel style={font=\color{white!15!black}},
ylabel={Decoding success rate},
axis background/.style={fill=white},
grid=both, 
major grid style={line width=.2pt,draw=gray!50},
minor grid style={line width=.1pt,draw=gray!20}
]
\addplot[ybar, bar width=5, fill=mycolor1, draw=mycolor1, area legend] table[row sep=crcr] {%
0   0.3824\\
1	0.2059\\
2	0.2647\\
3	0\\
4	0\\
5	0.0294\\
6	0.0882\\
7	0.0294\\
};
\addplot[forget plot, color=white!15!black] table[row sep=crcr] {%
-0.2	0\\
7.5	0\\
};
\end{axis}

\begin{axis}[%
width=5.833in,
height=4.375in,
at={(0in,0in)},
scale only axis,
xmin=0,
xmax=1,
ymin=0,
ymax=1,
axis line style={draw=none},
ticks=none,
axis x line*=bottom,
axis y line*=left
]
\end{axis}
\end{tikzpicture}%
        }
        \caption{\ac{MSA}}
    \end{subfigure}
    \hfill
    \begin{subfigure}[b]{0.48\textwidth}
        \centering
        \resizebox{\textwidth}{!}{
        \definecolor{mycolor1}{rgb}{0.00000,0.44700,0.74100}%
\begin{tikzpicture}

\begin{axis}[%
width=4.521in,
height=3.566in,
at={(0.758in,0.481in)},
scale only axis,
bar shift auto,
xmin=-0.2,
xmax=7.5,
xlabel style={font=\color{white!15!black}},
xlabel={$E_b$/$N_0$ [dB]},
ymin=0,
ymax=1,
ylabel style={font=\color{white!15!black}},
ylabel={Decoding success rate},
axis background/.style={fill=white},
grid=both, 
major grid style={line width=.2pt,draw=gray!50},
minor grid style={line width=.1pt,draw=gray!20}
]
\addplot[ybar, bar width=5, fill=mycolor1, draw=mycolor1, area legend] table[row sep=crcr] {%
0   0.3688\\
1	0.3812\\
2	0.1281\\
3	0.0437\\
4	0.0156\\
5	0.0625\\
6	0\\
7	0\\
};
\addplot[forget plot, color=white!15!black] table[row sep=crcr] {%
-0.2	0\\
7.5	0\\
};
\end{axis}

\begin{axis}[%
width=5.833in,
height=4.375in,
at={(0in,0in)},
scale only axis,
xmin=0,
xmax=1,
ymin=0,
ymax=1,
axis line style={draw=none},
ticks=none,
axis x line*=bottom,
axis y line*=left
]
\end{axis}
\end{tikzpicture}%
        }
        \caption{\ac{NMSA}}
    \end{subfigure}
    \caption{Decoding success rate for different $E_b$/$N_0$ values in the non-randomized case}
    \label{fig:ist_freq_notrand}
\end{figure}
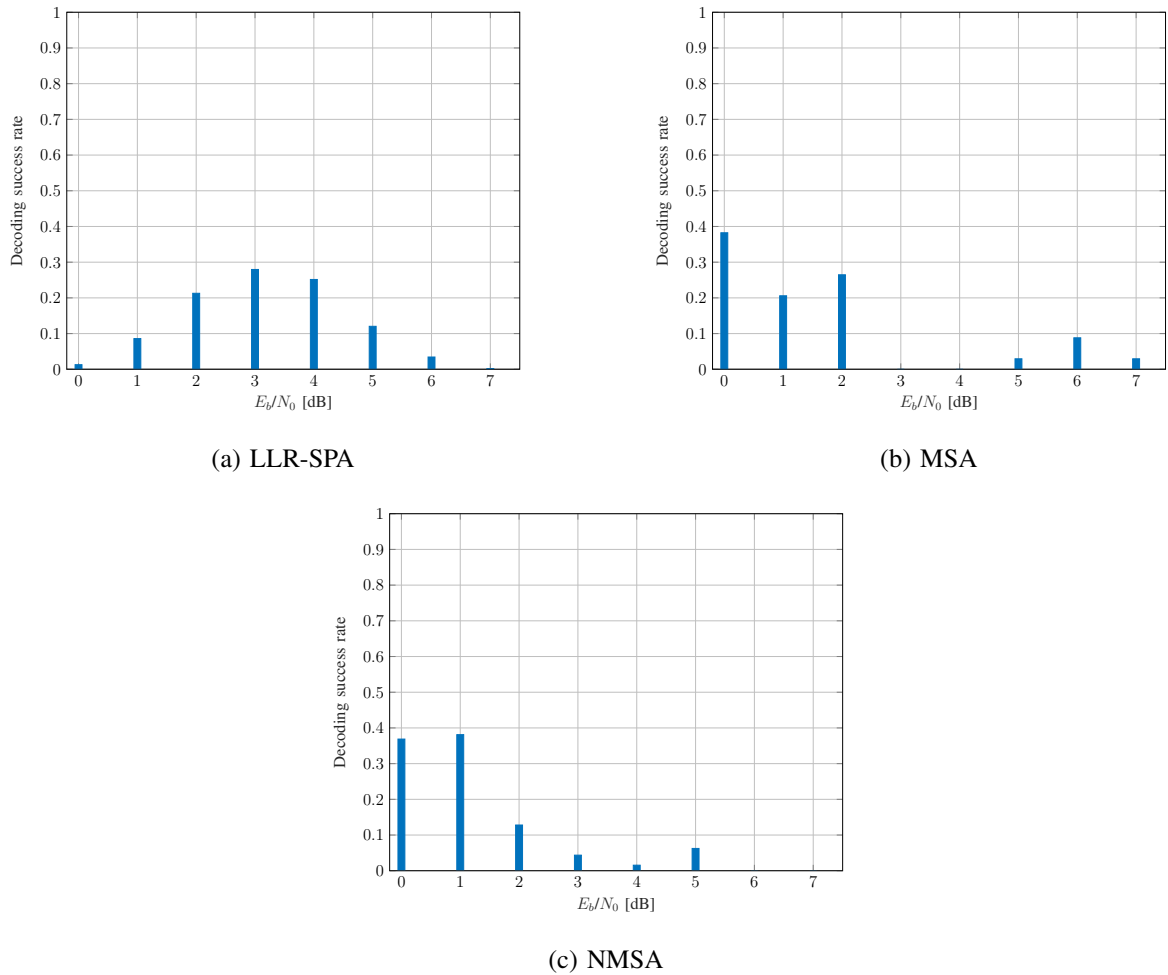

From Fig.s  \ref{fig:subspa} and \ref{fig:subspader} we observe that most of the decoding successes ($80\%$), causing a misinterpretation of the \ac{TS}, for the \ac{LLR-SPA} decoder, occur for $\frac{E_b}{N_0}$ between $2$ and $6$ dB, both included. This explains the results shown in Fig. \ref{fig:nat_notrand}; in fact, for the mentioned values of $\frac{E_b}{N_0}$, most decoding successes occur, leading to higher values of $P_{\mathrm{nat}}$ compared to other $\frac{E_b}{N_0}$ values. For the other decoders, the aforementioned effect is less pronounced. In particular, most of the decoding successes for the \ac{MSA} and the \ac{NMSA}-based decoders, in the de-randomized case, occur for small values of $E_b/N_0$. 

\section*{Acknowledgment}

The authors would like to thank Dr. Andrea Modenini for his precious insights.

The conclusions reported in this paper are the opinion of the authors and do not represent the official position of the European Space Agency.

\bibliographystyle{IEEEtran}
\bibliography{bibliography}

\end{document}